 \documentclass[final,5p,times,twocolumn,compress]{elsarticle}

\usepackage{amssymb}
\usepackage{amsmath}
\usepackage{lipsum}
\usepackage{color}

\journal{Physics Letters A}

\begin{document}

\begin{frontmatter}

\title{Dynamically stable two-mode squeezing in cavity optomechanics}
\author[first]{Chen Wang}
\author[first]{Shi-fan Qi\corref{cor1}}
\cortext[cor1]{Corresponding Author}
\affiliation[first]{organization={College of Physics and Hebei Key Laboratory of Photophysics Research and Application, Hebei Normal University},
         city={Shijiazhuang},
         postcode={050024}, 
         country={China}}
\ead{qishifan@hebtu.edu.cn}
            
\begin{abstract}
In this work, we propose a two-mode squeezing generation scheme in a hybrid three-mode cavity optomechanical system, where a mechanical resonator couples to two microwave (or optical) photon modes. By applying modulated strong drives, we derive an effective Hamiltonian that describes mechanically mediated two-photon squeezing, which is validated by diagonalizing the system’s transition matrix in the Heisenberg picture. Our analysis reveals that stable two-mode squeezing can be achieved by optimizing the squeezing operator even in unsteady-state dynamics, with the squeezing level exceeding the maximum achievable under system stability conditions while maintaining the anti-squeezing at a proper level within a suitable time interval. Furthermore, we show that our protocol is robust against systematic errors in both driving intensity and frequency, as well as against thermal Markovian noises. Our work presents an extendable approach for generating two-mode squeezed states between indirectly coupled bosonic modes.
\end{abstract}


\end{frontmatter}

\section{Introduction}
\label{introduction}
Quantum entanglement~\cite{quanentangle} plays a key role in quantum technologies, including quantum computing~\cite{quantumcomputing}, quantum communication~\cite{quantumcommunication}, and quantum sensing~\cite{quantumsense}. Many quantum platforms~\cite{cavityopto,cqed,Schrodingerstate,superconduct,magnoncavity,magnonqubit,rydberg} as well as protocols~\cite{mppentang,ghzstate,noon,atom,High,Cvoptical,Delocalization,entangledetection,ppnvcenter,Enhancementoem,WODEDO2025108364} to prepare and measure entangled states have therefore been intensively pursued for a long time and are still under active investigation. Among diverse entangled states, two-mode squeezed states (TMSS) are crucial in quantum computation~\cite{quancomcon}, information~\cite{quaninfcon,Gauinf}, teleportation~\cite{quantelep}, and metrology~\cite{quantmetro}. Various protocols have been proposed to generate TMSS with high squeezing level (SL)~\cite{nonlocalryd,kerrmagnon}, including mixing two single-mode squeezed states at a beam splitter~\cite{convariable} and employing a spontaneous parametric down-conversion process~\cite{twocolor,twinlaser,stablethreshold}. For the optical field, a non-degenerate optical parametric amplifier is often used to generate TMSS~\cite{quanphasenpo,realEPR,observasquee}. Recently, TMSS has been well established experimentally in various platforms, such as thermal gases~\cite{macroscopic}, ultra-cold atoms~\cite{homodyne,strongobserspin,probespin,spinBES,multisqueeatomic}, atomic mechanical oscillators~\cite{Atomicoscill}, spin ensembles~\cite{phasemeasure,manipulatinggrowth,powerlawspin}, antiferromagnetic magnons~\cite{antiferromagnetic}, and superconducting circuits~\cite{travelingwave,surfacephonon}. 

The cavity optomechanical system~\cite{cavityopto} provides an alternative and promising avenue to create optical~\cite{reservoiroptome,robustopto,doubleoptosys} and mechanical~\cite{twomechansquee,thermalsquee,Optoacoustic,amptmsnanomechanical} TMSS, due to its high controllability and flexibility. Two primary strategies are commonly employed to generate TMSS in this system. One approach involves constructing a two-mode squeezing interaction between the target modes~\cite{entanglemacro,probeentangle}, while the other uses reservoir engineering to tailor the target modes dissipatively into TMSS~\cite{reservoiroptome,robustopto,Entangling,nonreciprocal,generobust,Xie2023,synergizing}. The two-mode squeezing coupling naturally leads to entanglement without reservoir engineering. However, under the constraint of system stability conditions, the SL cannot go beyond $3$ dB below the vacuum limit~\cite{stablethreshold}. In contrast, the reservoir-engineering scheme ensures the system's stability and theoretically allows the SL to exceed the threshold. Nevertheless, due to the generally lower decay rate of phonons compared to photons~\cite{cavityopto,cohermechanical,entmechanical,squeemech,Superposition,coherentcoupling,active,map}, we find that the SL remains suboptimal and it is challenging to surpass the $3$ dB upper bound, even at absolute zero temperature. Consequently, beyond steady-state restrictions, some recent studies have turned to the dynamic generation of entanglement~\cite{Chen:17,entdynhightem} and the realization of other applications~\cite{Xie:20} in optomechanical systems. Moreover, squeezed states are capable of surpassing the standard quantum limit in precision sensing~\cite{SCHNABEL20171,Frequencyest}. In time-limited metrology, unsteady dynamical regimes may facilitate a faster generation of squeezing compared with steady-state schemes.

In this work, we adopt a similar dynamical perspective to generate TMSS in a three-mode optomechanical system, consisting of two target photon modes to be entangled, each coupled to an auxiliary phonon mode. The squeezing generation process is governed by an effective Hamiltonian for two-photon squeezing (TPS) coupling, which is validated by diagonalization of the system's transition matrix in the Heisenberg picture. By analyzing the system's dynamics through the effective TPS Hamiltonian in Markovian environments, we find that an asymptotically stable TMSS, characterized by the squeezing operator variance that gradually converges to a constant over time, can be obtained in unsteady evolutions, where the system's covariance matrix (CM) elements diverge exponentially, displaying an enhanced SL exceeding the steady limit. Environmental noises modify the optimized operator of TPS while simultaneously asymptotically stabilizing TMSS, even if the TPS coupling exceeds the system stability threshold~\cite{stablethreshold}. Furthermore, numerical simulations show that our protocol can maintain the anti-two-mode-squeezing at an appropriate level within a suitable time interval and is robust against systematic errors in both driving intensity and frequency, as well as against thermal Markovian noises.

The rest of this work is organized as follows. In Sec.~\ref{Secmodel}, we present a three-mode optomechanical system and derive an effective TPS Hamiltonian assisted by the phonon. In Sec.~\ref{Sectms}, we phenomenologically analyze the generation process of the TMSS within the open-quantum-system framework. We find that the stable TPS can be obtained even beyond the system stability conditions. In Sec.~\ref{secanti}, we discuss the anti-two-mode-squeezing effect beyond the system-stability regime. Systematic errors arising from the driving-enhanced optomechanical coupling strengths and driving frequencies, as well as the effects of thermal Markovian noises, are analyzed in Secs.~\ref{secsyserror} and~\ref{secthermal}, respectively. Finally, we discuss the experimental feasibility and summarize the work in Sec.~\ref{Secexpercon}.

\section{Model and the effective Hamiltonian}\label{Secmodel}
\begin{figure}[htbp]
\centering
\includegraphics[width=0.45\textwidth]{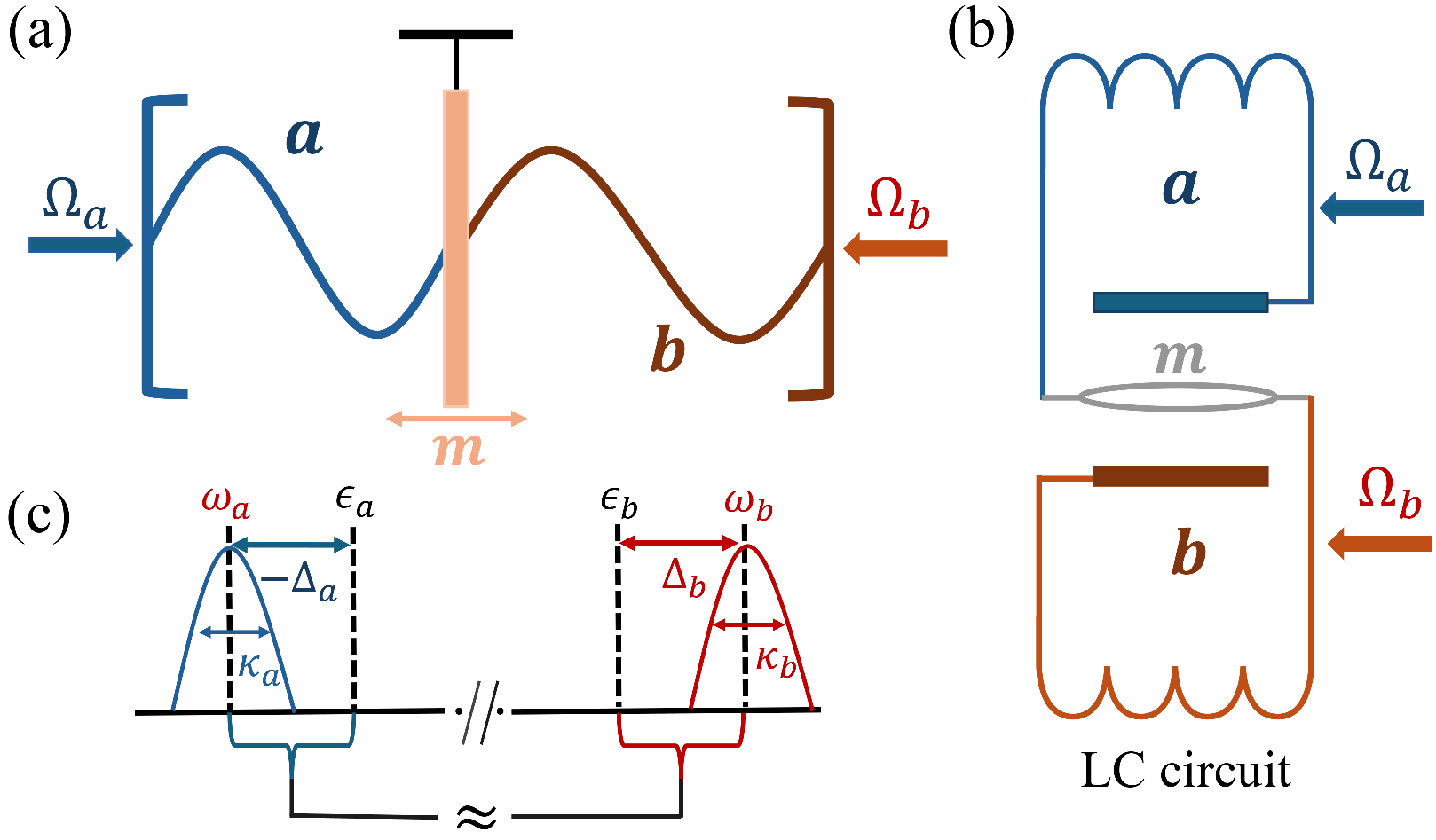}
\caption{Schematic diagram of the hybrid three-mode cavity optomechanical system. (a) A mechanical interface acts as an intermediate mode $m$, coupling with the optical cavity $a$ and $b$. (b) A mechanical resonator $m$ is capacitively coupled to two superconducting microwave resonators, $a$ and $b$. The photon modes $a$ and $b$ are driven by strong fields $\Omega_a$ and $\Omega_b$, respectively. (c) The frequencies and linewidths of the system are adopted to generate TMSS.}\label{diagram}
\end{figure}
Consider a hybrid three-mode optomechanical system as shown in Fig.~\ref{diagram}, which is composed of a mechanical oscillator and two optical cavity modes [see Fig.~\ref{diagram} (a)], or a mechanical oscillator and two superconducting microwave resonators [see Fig.~\ref{diagram} (b)]. The full system Hamiltonian ($\hbar\equiv1$) can be described as~\cite{cavityopto,reservoiroptome}
\begin{equation}\label{Hamori}
	H=\omega_mm^\dag m+\sum_{o=a,b}\omega_oo^\dag o+g_oo^\dag o(m+m^\dag)+\Omega_o (oe^{i\epsilon_ot}+o^\dag e^{-i\epsilon_ot}),
\end{equation}
where $a(a^\dag)$, $b(b^\dag)$, and $m(m^\dag)$ are the annihilation (creation) operators of two photon modes and phonon mode, with transition frequencies $\omega_a$, $\omega_b$, and $\omega_m$, respectively. $g_a$ ($g_b)$ is the single-excitation photon-phonon coupling strength between photon mode $a$ ($b$) and phonon mode $m$, which can be compensated by a strong drive. The last term describes the external driving Hamiltonian, where $\Omega_o$ is the Rabi frequency and $\epsilon_o$ is the driving frequency of mode $o$, $o=a,b$.

Under appropriate strong driving conditions and following the standard linearization approach~\cite{cavityopto}, the full system Hamiltonian turns out to be
\begin{equation}\label{Hamline}
\begin{aligned}
	H_{\rm lin}&=H_0+V,\quad H_0=\Delta_aa^\dag a+\Delta_bb^\dag b+\omega_mm^\dag m,\\
	V&=g(e^{-i\theta_a}a+e^{i\theta_a}a^\dag)(m+m^\dag)+G(e^{-i\theta_b}b+e^{i\theta_b}b^\dag)(m+m^\dag),
\end{aligned}
\end{equation}
where $\Delta_a=\omega_a-\epsilon_a$ and $\Delta_b=\omega_b-\epsilon_b$ are the detunings of mode $a$ and $b$, respectively. $g$ and $G$ are the driving-enhanced optomechanical coupling strengths, $\theta_a$ and $\theta_b$ are the corresponding phases. The details can be found in~\ref{appaline}.

In previous works using reservoir engineering method~\cite{reservoiroptome}, the parameters were set as $\Delta_a=\omega_m$, $\Delta_b=-\omega_m$, and $g<G$, to obtain the two-mode squeezing (quantum entanglement) between modes $a$ and $b$. These conditions ensure the system's stability and theoretically enable the SL to exceed the $3$ dB. However, due to the generally lower decay rate of phonon compared to the photon modes~\cite{cohermechanical,entmechanical,squeemech,Superposition,coherentcoupling,active,map}, even at zero temperature, the SL remains suboptimal and is difficult to surpass the upper bound of $3$ dB. More details and analysis can be seen in~\ref{appbtmsssteady}. To overcome this limitation, we focus on generating the TMSS by constructing an effective TPS Hamiltonian within the framework of system instability.

At the large detuning regime, i.e., $|\Delta_a-\omega_m|, |\Delta_b-\omega_m|\gg g, G$, and under the near-resonant condition $\Delta_a=-\Delta_b+\delta$, an effective Hamiltonian describing the TPS can be extracted by a perturbation theory~\cite{james}. The effective Hamiltonian is found to be
\begin{equation}\label{Heff}
	H_{\rm eff}=g_{\rm
		eff}(e^{-i\theta}ab+e^{i\theta}a^\dag b^\dag),
\end{equation}
where $\theta=\theta_a+\theta_b$. The effective coupling strength and the energy shift are
\begin{equation}\label{geffdelta}
	g_{\rm eff}=\frac{2\omega_mgG}{\Delta^2_b-\omega^2_m},\quad \delta=\frac{2\omega_m(g^2+G^2)}{\omega^2_m-\Delta^2_b},
\end{equation}
respectively. The derivation details are presented in~\ref{appcHeff}. The effective Hamiltonian in Eq.~(\ref{Heff}) can naturally generate the TMSS without reservoir engineering~\cite{stablethreshold}.

The effective TPS Hamiltonian does not conserve excitation, making it challenging to validate the effective Hamiltonian using previous methods~\cite{intermagnon,oneexcite,nonlinear}, i.e., a standard numerical diagonalization of the system Hamiltonian in a truncated Hilbert space. To address this difficulty, we introduce an interesting approach involving diagonalization of the transition matrix of the whole system, which enables the evaluation of the effective two-mode squeezing Hamiltonian induced by the virtual process. The details can be seen in~\ref{appcHeff}. In~\ref{appcHeff}, we benchmark the validity ranges of the coupling strengths $g$ and $G$, as well as the detunings $\Delta_b$, by comparing the analytical predictions for $g_{\rm eff}$ and $\delta$ derived from the effective Hamiltonian with numerical simulations based on the full system's transition matrix. The results identify an approximate parameter regime in which the effective Hamiltonian remains both valid and well-performing, namely, $0.1\omega_m\le g\le0.2\omega_m, 8g\le(\Delta_b-\omega_m)\le10g$. Within this region, the effective coupling strength satisfies $g_{\rm eff}\gtrsim0.01\omega_m$, while both the absolute and relative errors remain small.

\section{Two-photon squeezing}\label{Sectms}
Using the effective Hamiltonian in Eq.~(\ref{Heff}), one can naturally generate the TMSS between two photon modes $a$ and $b$. Within the open-quantum-system framework, this section analyzes the system's dynamics and elucidates the generation mechanism of stable TMSS under system instability conditions. Under the standard Markovian assumptions, the dynamics of the quantum system are governed by the quantum Langevin equation, written in a matrix form
\begin{equation}\label{qleeff}
	\dot{u}^{\rm eff}(t)=A_{\rm eff}u^{\rm eff}(t)+n^{\rm eff}(t),
\end{equation}
where $u^{\rm eff}(t)=[X_a(t), Y_a(t), X_b(t), Y_b(t)]^T$ with $X_o=(e^{-i\theta_o}o+e^{i\theta_o}o^\dag)/\sqrt{2}$, $Y_o=(e^{-i\theta_o}o-e^{i\theta_o}o^\dag)/i\sqrt{2}$, $o=a,b$. The transition matrix $A_{\rm eff}$ is
\begin{equation}\label{aeff}
A_{\rm eff}=-\begin{bmatrix}\kappa_a & 0 &0 &g_{\rm eff}\\
0 &\kappa_a &g_{\rm eff} &0\\
0 & g_{\rm eff} &\kappa_b &0\\
g_{\rm eff} & 0 & 0 &\kappa_b\end{bmatrix},
\end{equation} 
where $n^{\rm eff}(t)=[\sqrt{2\kappa_a}X_a^{in}(t),\sqrt{2\kappa_a}Y_a^{in}(t),\sqrt{2\kappa_b}X_b^{in}(t),\sqrt{2\kappa_b}Y_b^{in}(t)]^T$ is the vector of Gaussian noise operators, and $X^{in}_o=(e^{-i\theta_o}o_{in}+e^{i\theta_o}o_{in}^\dag)/\sqrt{2}$,$Y^{in}_o=(e^{-i\theta_o}o_{in}-e^{i\theta_o}o_{in}^\dag)/i\sqrt{2}$. $\kappa_a$ and $\kappa_b$ are the decay rates of the modes $a$ and $b$, respectively. $o_{in}$ is the input noise operators for the mode $o$, which is characterized by the covariance functions: $\langle o^\dag_{in}(t)o_{in}(t')\rangle=N_o\delta(t-t')$ and $\langle o_{in}(t)o^\dag_{in}(t')\rangle=[N_o+1]\delta(t-t')$, under the Markovian approximation. $N_o$ is the thermal occupation number of mode $o$ at equilibrium state and $N_o=[\exp(\hbar\omega_o/k_BT)-1]^{-1}$.

At the initial time, we assume the microwave mode $a$ and optical mode $b$ are both in thermal equilibrium states. Then, the system state evolves as a Gaussian state, due to the above-linearized dynamics in Eq.~(\ref{qleeff}), which can be completely characterized by a $4\times 4$ CM $V^{\rm eff}(t)$. The dynamics of the CM $V^{\rm eff}(t)$ satisfies
\begin{equation}\label{cmeff}
	\dot{V}^{\rm eff}(t)=A_{\rm eff}V^{\rm eff}(t)+V^{\rm eff}(t)A^T_{\rm eff}+D^{\rm eff}.
\end{equation}
The elements of $V^{\rm eff}(t)$ are defined as
\begin{equation}\label{Veffdefin}
	V^{\rm eff}_{ij}(t)=\frac{\langle u^{\rm eff}_i(t)u^{\rm eff}_j(t)+u^{\rm eff}_j(t)u^{\rm eff}_i(t)\rangle}{2}-\langle u^{\rm eff}_i(t)\rangle\langle u^{\rm eff}_j(t)\rangle,
\end{equation}
where $u^{\rm eff}_i(t)$ is the $i$ term of $u^{\rm eff}(t)$. $D^{\rm eff}$ is the diffusion matrix, which is diagonal with elements given by $D^{\rm eff}_{11}=D^{\rm eff}_{22}=\kappa_a(2N_a+1)$ and $D^{\rm eff}_{33}=D^{\rm eff}_{44}=\kappa_b(2N_b+1)$.
This matrix is defined through $D^{\rm eff}_{ij}(t)=\langle n^{\rm eff}_i(t)n^{\rm eff}_j(t)+n^{\rm eff}_j(t)n^{\rm eff}_i(t)\rangle/2$. The CM elements can be analytically calculated, with explicit expressions provided in~\ref{appcm}, Eq.~(\ref{Veffelement}). In the system stability condition, the CM is invariant, i.e., $\dot{V}^{\rm eff}=0$ in Eq.~(\ref{cmeff}), which requires $g^2_{\rm eff}<\kappa_a\kappa_b$. In regimes of system dynamical instability, the CM diverges.

Beyond the system stability region, $g^2_{\rm eff}>\kappa_a\kappa_b$, we demonstrate that stable, rather than merely transient, two-mode squeezing can emerge, by optimizing the two-mode squeezing quadrature operator. The quadrature operator is given by
\begin{equation}\label{quadratureX}
	X=\cos\tilde{\phi}X_a+\sin\tilde{\phi}Y_b,
\end{equation}
and the associated variance becomes
\begin{equation}\label{Deltaxphit}
	\Delta X(t)=2C_-e^{-(\Omega+\kappa_a+\kappa_b)t}+\frac{\kappa_++\cos(2\tilde{\phi})\kappa_-}{2(\Omega+\kappa_a+\kappa_b)},
\end{equation}
where
\begin{equation}\label{Omegphi}
\begin{aligned}
&\tan(2\tilde{\phi})=\frac{2g_{\rm eff}}{\kappa_a-\kappa_b},\quad \Omega=\sqrt{4g^2_{\rm eff}+(\kappa_a-\kappa_b)^2},\\
&\kappa_\pm=\kappa_a(2N_a+1)\pm\kappa_b(2N_b+1),\quad N_\pm=N_a\pm N_b,\\
&C_-=\left[\frac{N_++1+\cos(2\tilde{\phi})N_-}{4}-\frac{\kappa_++\cos(2\tilde{\phi})\kappa_-}{4(\Omega+\kappa_a+\kappa_b)}\right].
\end{aligned}
\end{equation}
More details can be seen in~\ref{appcm}.

The condition $\Delta X(t)<0.5$ indicates the emergence of two-mode squeezing, with smaller values of $\Delta X(t)$ corresponding to higher squeezing. Equation~(\ref{Deltaxphit}) shows an asymptotically stable TPS over a long time evolution, i.e.,
\begin{equation}\label{Deltaxphiinfty}
	\Delta X(\infty)=\frac{\Omega\kappa_++(\kappa_a-\kappa_b)\kappa_-}{2\Omega(\Omega+\kappa_a+\kappa_b)}.
\end{equation}
Based on the definition of $\Omega$ in Eq.~(\ref{Omegphi}), $\Delta X(\infty)$ decreases monotonically with increasing $g_{\rm eff}$, indicating a corresponding enhancement of the TPS. Additionally, we define a sufficiently long period $\tau$ to approximate infinite time, which is given by $\tau=2\pi/(\Omega+\kappa_a+\kappa_b)$. It can be demonstrated that the difference $\Delta X(\tau)-\Delta X(\infty)=2C_-/e^{2\pi}$ is sufficiently small. We also simulate the SL to quantify the TPS, which in the decibel unit is defined by 
\begin{equation}\label{squzelevel}
	S=-10{\rm log}_{10}\left(\Delta X/\Delta X_{zp}\right),
\end{equation}
where $\Delta X_{zp}=0.5$ is the fluctuation in the zero-point level.

Moreover, we numerically calculate the whole system's dynamic using the full Hamiltonian $H_{\rm lin}$ in Eq.~(\ref{Hamline}) to confirm the above analytical results. Similar to Eq.~(\ref{cmeff}), the dynamic of the whole system CM $V(t)$ satisfies
\begin{equation}\label{wholedotV}
	\dot{V}(t)=AV(t)+V(t)A^T+D.
\end{equation}
The elements of $V(t)$ are given by
\begin{equation}\label{Vt}
	V_{ij}=\frac{\langle u_i(t)u_j(t)+u_j(t)u_i(t)\rangle}{2}-\langle u_i(t)\rangle\langle u_j(t)\rangle,
\end{equation}
where $u_i(t)$ is the $i$ term of $u(t)$ and $i=1,2,\cdots,6$. $u^T(t)=[u^{\rm eff}(t),X_m(t),Y_m(t)]$ is the vector of quadrature operators, where $u^{\rm eff}$ is defined in Eq.~(\ref{qleeff}), $X_m=(m+m^\dag)/\sqrt{2}$, and $Y_m=(m-m^\dag)/i\sqrt{2}$. $A=i\mathcal{L}-\tilde{A}$, where $\mathcal{L}$ is a matrix defined in Eq.~(\ref{matrix}) in~\ref{appcHeff}, and $\tilde{A}=Diag[\kappa_a,\kappa_a,\kappa_b,\kappa_b,\kappa_m,\kappa_m]$. $D$ is the matrix of noise covariance, where the diagonal elements are 
$D_{11}=D_{22}=\kappa_a(2N_a+1)$, $D_{33}=D_{44}=\kappa_b(2N_b+1)$, and $D_{55}=D_{66}=\kappa_m(2N_m+1)$, all non-diagonal elements are zero. Then, the dynamic of $\Delta X(t)$ can be numerically obtained by calculating the CM $V(t)$,
\begin{equation}\label{deltaxfull}
	\Delta X(t)=\cos^2\tilde{\phi} V_{11}(t)+\sin^2\tilde{\phi} V_{44}(t)+\sin(2\tilde{\phi})V_{14}(t).
\end{equation}
Here, the mechanical mode is initially assumed to be in the vacuum state, and its corresponding environment noises are zero-mean Gaussian noises.

\begin{figure}[t]
	\centering
	\includegraphics[width=0.235\textwidth]{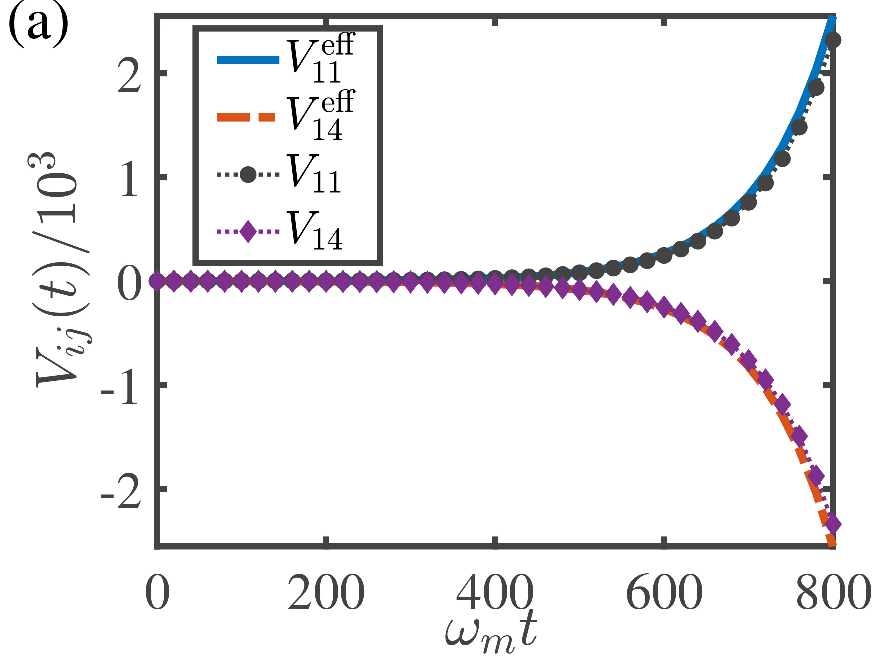}
	\includegraphics[width=0.235\textwidth]{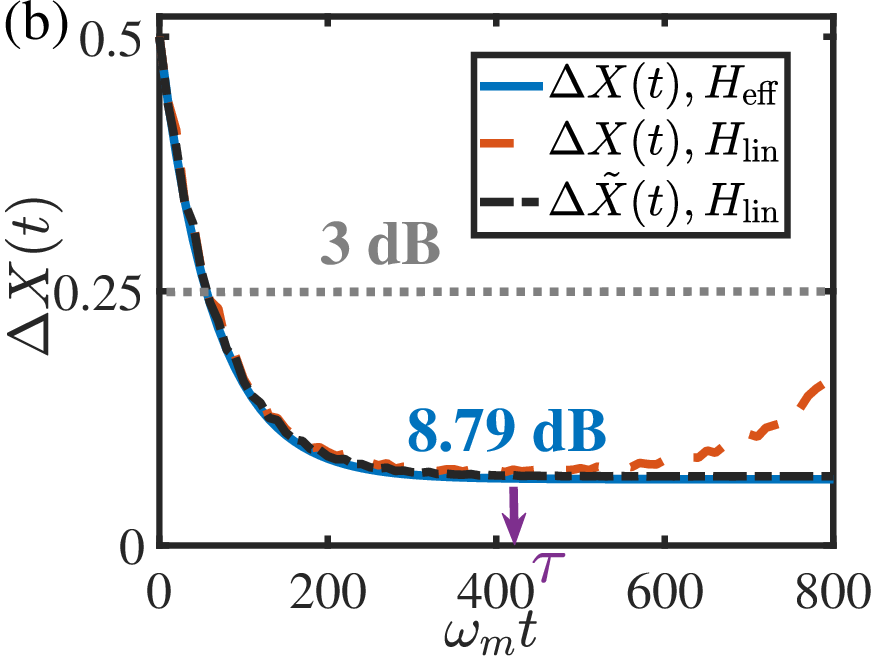}
	\caption{(a) Dynamics of the CM elements using the effective Hamiltonian~\eqref{Heff} or the full system Hamiltonian~\eqref{Hamline}. (b) Dynamics of the $\Delta X(t)$ and $\Delta\tilde{X}(t)$ with the effective Hamiltonian~\eqref{Heff} or the full system Hamiltonian~\eqref{Hamline}. Here, the parameters are set as $g=G=0.1\omega_m,\Delta_b=\omega_m+10g,\kappa_a=\kappa_b=10^{-3}\omega_m,\kappa_m=10^{-6}\omega_m$, and the thermal occupation numbers $N_a=N_b=0,N_m=10$. Moreover, all modes are initially assumed to be in vacuum states, i.e., $V(0)=I_6/2$, where $I_6$ is the $6\times 6$ identity matrix.}
	\label{vtsqueeze}
\end{figure}

In Fig.~\ref{vtsqueeze} (a), it can be observed that both of the CM elements $V_{11}$ (dark dotted line with circles) and $V_{14}$ (purple dotted line with diamonds) exhibit excellent agreement with the corresponding analytical results via the effective Hamiltonian within the time regime $\omega_mt\le 600$. In Fig.~\ref{vtsqueeze} (b), we plot $\Delta X(t)$ using the effective Hamiltonian in Eq.~(\ref{Heff}) by the blue solid line. It tends to stabilize a certain value of $0.066$ after prolonged evolution ($\omega_m t\ge250\approx 0.6\tau$), and the corresponding SL is approximately $8.79$ dB, which is greater than the upper bound $3$ dB in the system stability condition. However, for longer time evolution, it is found that the result about $\Delta X(t)$ using the effective Hamiltonian~(\ref{Heff}) differs obviously from that using the full Hamiltonian~(\ref{Hamline}) (red dashed line), although the CM elements remain in close agreement. In the system instability regime, the CM elements $V_{11}$ and $V_{14}$ exhibit exponential dependence on time, as shown in Fig.~\ref{vtsqueeze} (a). Over a long time, the absolute values of these elements rapidly increase. In this condition, even a slight deviation in the CM elements can significantly affect the variance $\Delta X(t)$, decreasing the SL instead of the expected enhancement. However, it is observed that the squeeze remains nearly stable within the approximate time interval $t\in[0.75\tau,1.25\tau]$.

Since the effective Hamiltonian fails to fully capture the whole system's evolution over a long time, the optimal squeezing operator in practice is not the theoretical prediction $X$ shown in Eq.~(\ref{quadratureX}). The genuine optimal operator $\tilde{X}$ is obtained by numerically minimizing the $\Delta\tilde{X}$ of the general quadrature operator
$(X_a\cos\phi_1+Y_a\sin\phi_1)\cos\phi_3+(X_b\cos\phi_2+Y_b\sin\phi_2)\sin\phi_3$, with respect to $\phi_1$, $\phi_2$, and $\phi_3$. It is demonstrated that the variance $\Delta\tilde{X}$ of the genuine optimal operator $\tilde{X}$ corresponds to the minimal eigenvalue of sub-CM $V_4=V(1:4,1:4)$, where $V$ is the full CM defined in Eq.~(\ref{wholedotV}). The variance $\Delta\tilde{X}(t)$, numerically calculated by the full system Hamiltonian in Eq.~(\ref{Hamline}), is presented as a black dash-dotted line in Fig.~\ref{vtsqueeze} (b). These results do match well with the theoretical predictions (blue solid line) and maintain consistency over time. The discrepancy between the results for $\Delta X$ and $\Delta\tilde{X}$ becomes increasingly pronounced over time. This behavior arises because the relevant CM elements are exponentially amplified in the unsteady-state regime, such that even a slight misalignment of the quadrature operator can lead to a substantial deviation in the observed SL. Consequently, there exists a favorable time window in which the squeezing remains appreciably strong while the requirement for angular precision is comparatively relaxed. We further discuss this favorable time window and the associated angular precision in Sec.~\ref{secanti}.

\begin{figure}[t]
	\centering
	\includegraphics[width=0.235\textwidth]{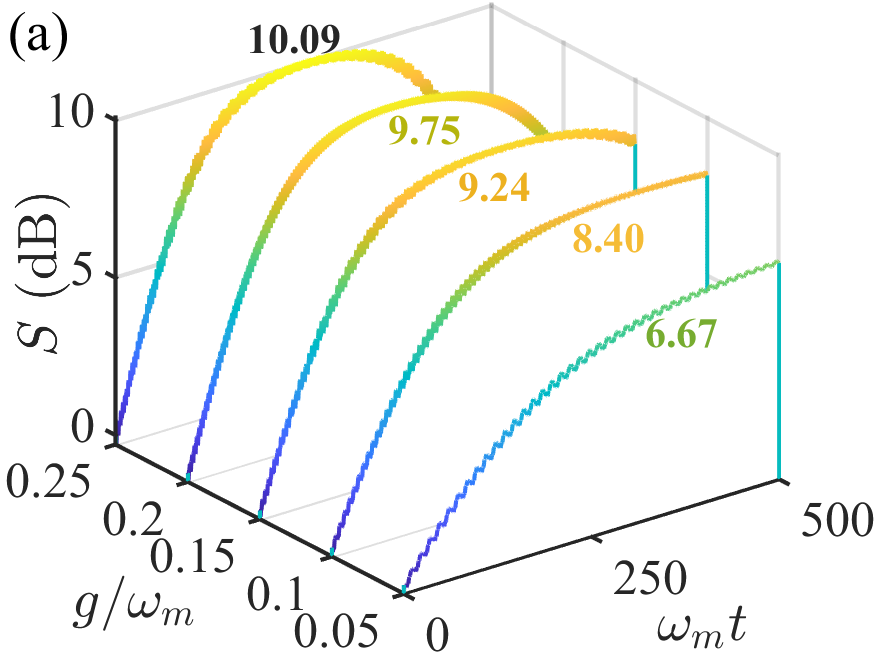}
	\includegraphics[width=0.235\textwidth]{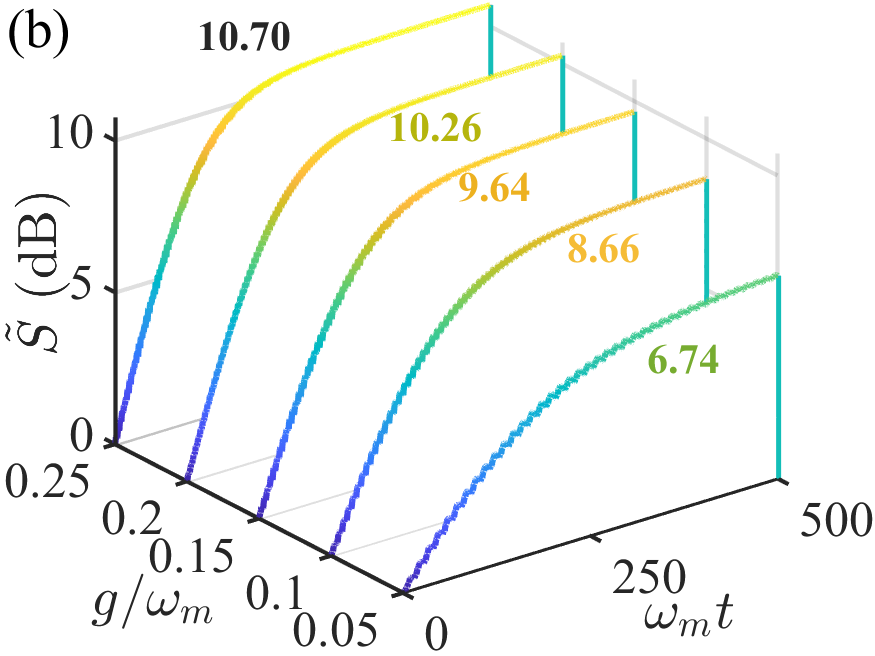}
	\includegraphics[width=0.235\textwidth]{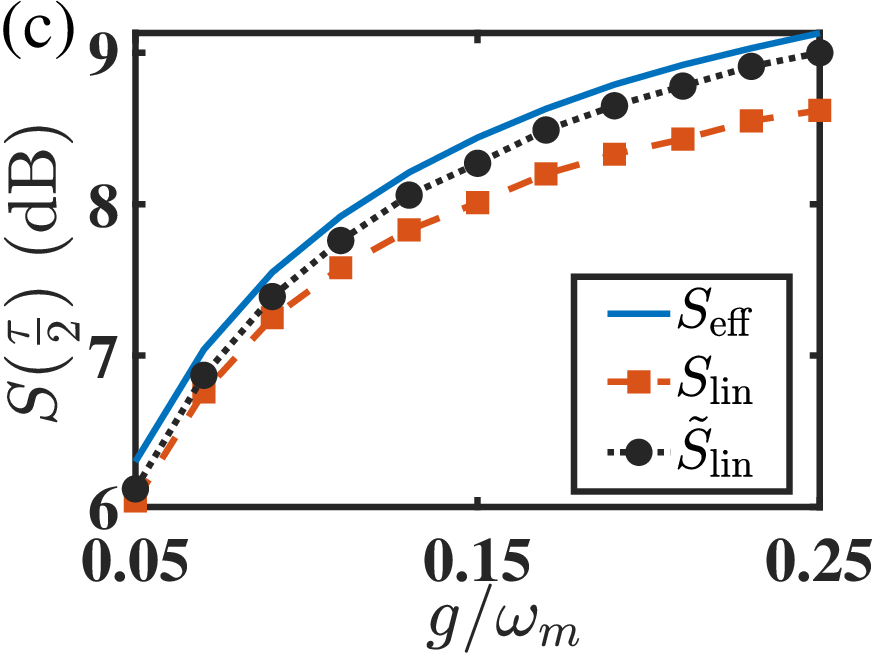}
	\includegraphics[width=0.235\textwidth]{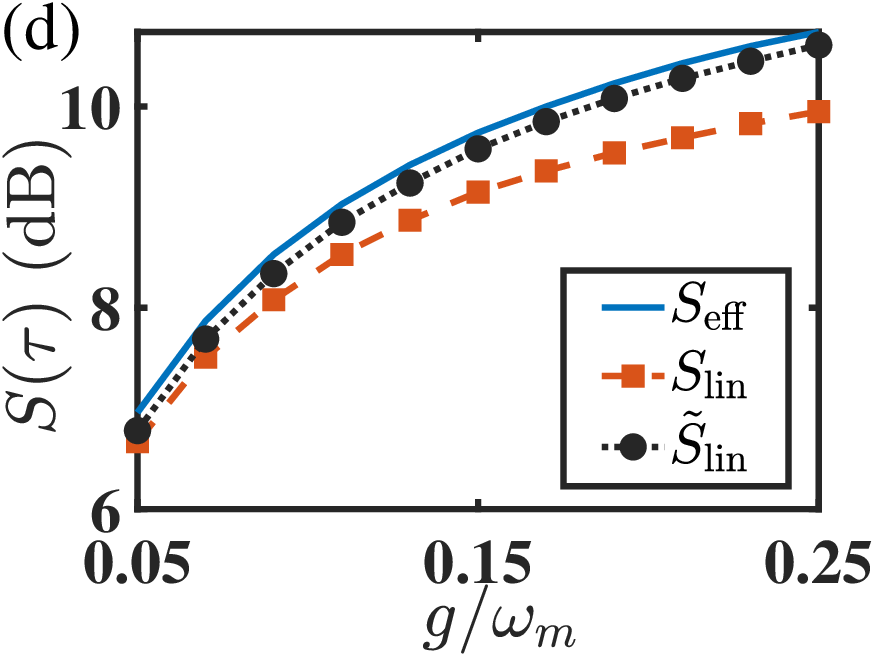}
	\includegraphics[width=0.235\textwidth]{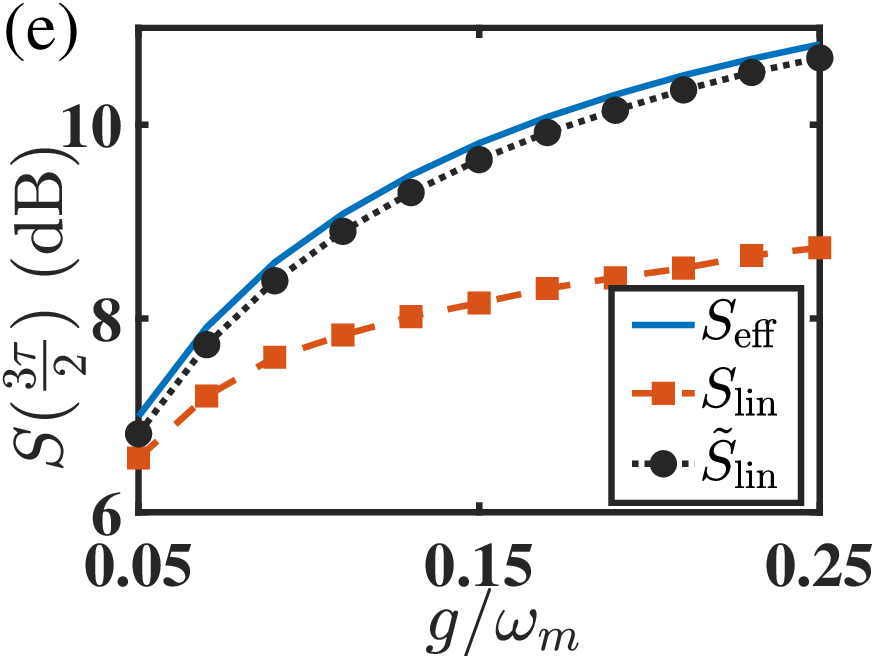}
	\includegraphics[width=0.235\textwidth]{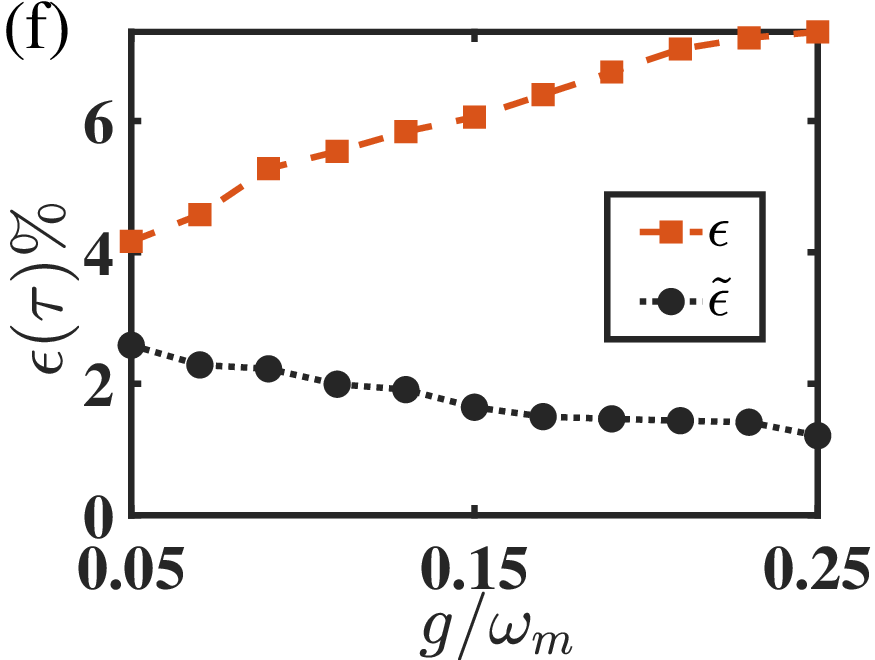}
	\caption{(a) Time evolution of the SL $S$ for operator $X$~\eqref{quadratureX} under the full system Hamiltonian~\eqref{Hamline} at varying coupling strengths. (b) Time evolution of the SL $\tilde{S}$ for numerically optimized quadrature operator $\tilde{X}$ with the full system Hamiltonian~\eqref{Hamline} at different coupling strengths. [(c), (d), and (e)] Comparison of the analytical SL $S_{\rm eff}$ with the numerically calculated results $S_{\rm lin}$ and $\tilde{S}_{\rm lin}$ under different normalized coupling strengths $g/\omega_m$. (f) The relative errors of SL, $\epsilon$ and $\tilde{\epsilon}$, as functions of $g/\omega_m$. All other parameters and initial conditions are the same as those in Fig.~\ref{vtsqueeze}.}
	\label{squeeze}
\end{figure}

In Figs.~\ref{squeeze} (a) and (b), we plot the dynamics of the SL $S$ [as defined in Eq.~(\ref{squzelevel})] and $\tilde{S}$ using the full system Hamiltonian~(\ref{Hamline}) under various coupling strengths, respectively. Here, the quantity $\tilde{S}$ represents the SL of the genuine optimal quadrature $\tilde{X}$, which is defined as $\tilde{S}=-10{\rm log}_{10}\left(\Delta \tilde{X}/\Delta X_{zp}\right)$. The numerical annotations in the figure represent the maximum SL attained during the evolution period. From Figs.~\ref{squeeze} (a) and (b), one can conclude that the SL is enhanced by increasing the coupling strength, which is consistent with the analytical results. Furthermore, as shown in Fig.~\ref{squeeze} (a), the duration of stable squeezing for $S$ decreases as the coupling strength increases. For $g=0.05\omega_m$ and $g=0.1\omega_m$, the SL has no significant degradation during the time interval ($\omega_mt\le 500$). However, when $g=0.2\omega_m$ and $g=0.25\omega_m$, a notable decline in the SL is observed at $\omega_mt\approx350$ and $\omega_mt\approx300$, respectively. In contrast, the SL $\tilde{S}$ of the numerically optimized quadrature $\tilde{X}$ maintains asymptotic stable upon attaining a plateau, independent of the coupling strength.

In Figs.~\ref{squeeze} (c), (d), and (e), the SL $S_{\rm eff}$ for quadrature operator $X$~(\ref{quadratureX}) with the effective Hamiltonian~(\ref{Heff}), the SL $S_{\rm lin}$ for $X$~(\ref{quadratureX}) with the full system Hamiltonian~(\ref{Hamline}), and the SL $\tilde{S}_{\rm lin}$ for the numerically optimized operator $\tilde{X}$ with full system Hamiltonian~(\ref{Hamline}), are shown with blue solid line, red dashed line with squares, and black dotted line with circles, respectively. Panels (c), (d), and (e) display the SL at specific moments $\tau/2$, $\tau$, and $3\tau/2$, respectively. Comparison of the results in Figs.~\ref{squeeze} (c), (d), and (e) reveals that, regardless of the coupling strength, the SL $S_{\rm eff}$ and $\tilde{S}_{\rm lin}$ at time $\tau$ exhibit only a slight difference compared to those at $3\tau/2$, while being $1-2$ units larger than the corresponding values at $\tau/2$. The two-mode squeezing has become stable at time $\tau$. In contrast, the SL $S_{\rm lin}$ at $3\tau/2$ is consistently smaller than those at $\tau$, with the magnitude of this difference growing as the coupling strength $g$ increases. This implies that, when selecting $X$ as the quadrature operator, while the SL enhances with stronger coupling, the duration of high SL decreases proportionally. These observations are consistent with the theoretical predictions and the numerical results shown in Figs.~\ref{squeeze} (a) and (b).

We introduce the relative errors of SL $S_{\rm lin}$ and $\tilde{S}_{\rm lin}$ to more clearly demonstrate the validity of our protocol. The relative errors are defined as
\begin{equation}\label{epsilon}
	\epsilon(t)=\left|\frac{S_{\rm lin}(t)-S_{\rm eff}(t)}{S_{\rm eff}(t)}\right|, \tilde{\epsilon}(t)=\left|\frac{\tilde{S}_{\rm lin}(t)-S_{\rm eff}(t)}{S_{\rm eff}(t)}\right|,
\end{equation}
respectively, which numerical results at characteristic moment $\tau$ are presented in Fig.~\ref{squeeze} (f). As the coupling strength $g$ increases, the relative error $\epsilon(\tau)$ associated with SL $S_{\rm lin}(\tau)$ roughly tends to increase, whereas the relative error $\tilde{\epsilon}(\tau)$ corresponding to the optimal operator SL $\tilde{S}_{\rm lin}(\tau)$ shows a gradual decrease. Furthermore, the $\tilde{\epsilon}(\tau)$ remains significantly smaller than $\epsilon(\tau)$ for all coupling strengths. Specifically, for $g\ge0.15\omega_m$, $\tilde{\epsilon}(\tau)\le 0.02$ while $\epsilon(\tau)\ge0.06$. The relative error $\tilde{\epsilon}$ closely follows the trend of the relative error of effective coupling, $\sigma$, shown in Fig.~\ref{effgdelta} (f) in~\ref{appcHeff}, while the $\epsilon$ shows a distinctly opposite trend. It indicates that, although the effective Hamiltonian can approximately capture the whole system dynamics, the squeezing operator $X$ in Eq.~(\ref{quadratureX}) derived from this framework deviates slightly from the genuine optimal operator.

\section{Performance analysis}
\subsection{Anti-two-mode-squeezing effect}\label{secanti}
\begin{figure}[ht]
	\centering
	\includegraphics[width=0.235\textwidth]{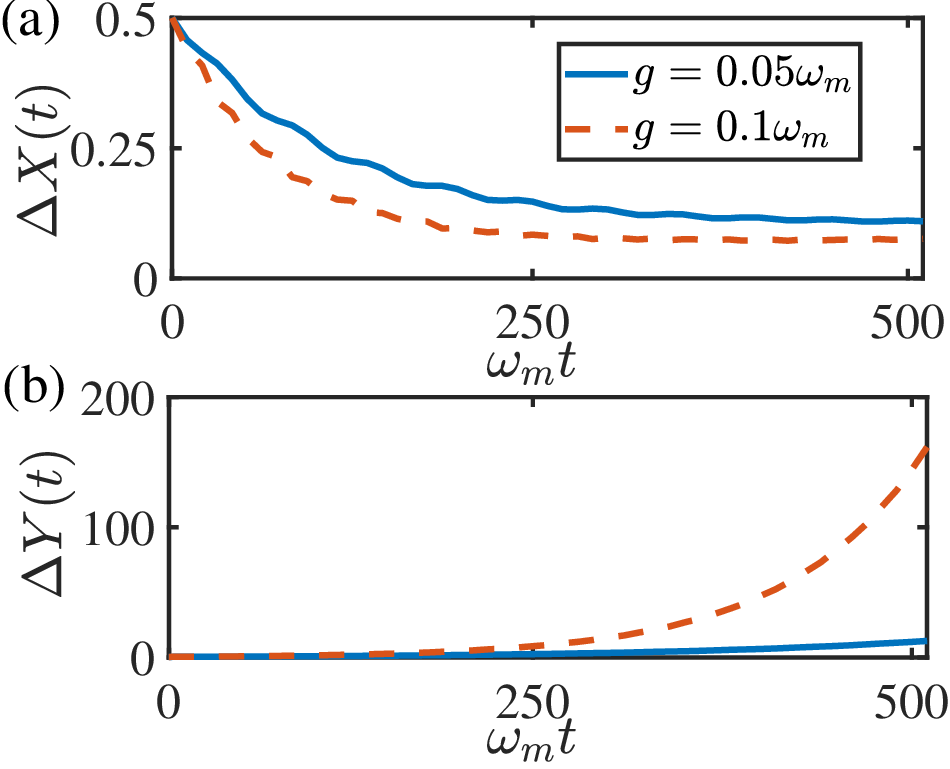}
	\includegraphics[width=0.235\textwidth]{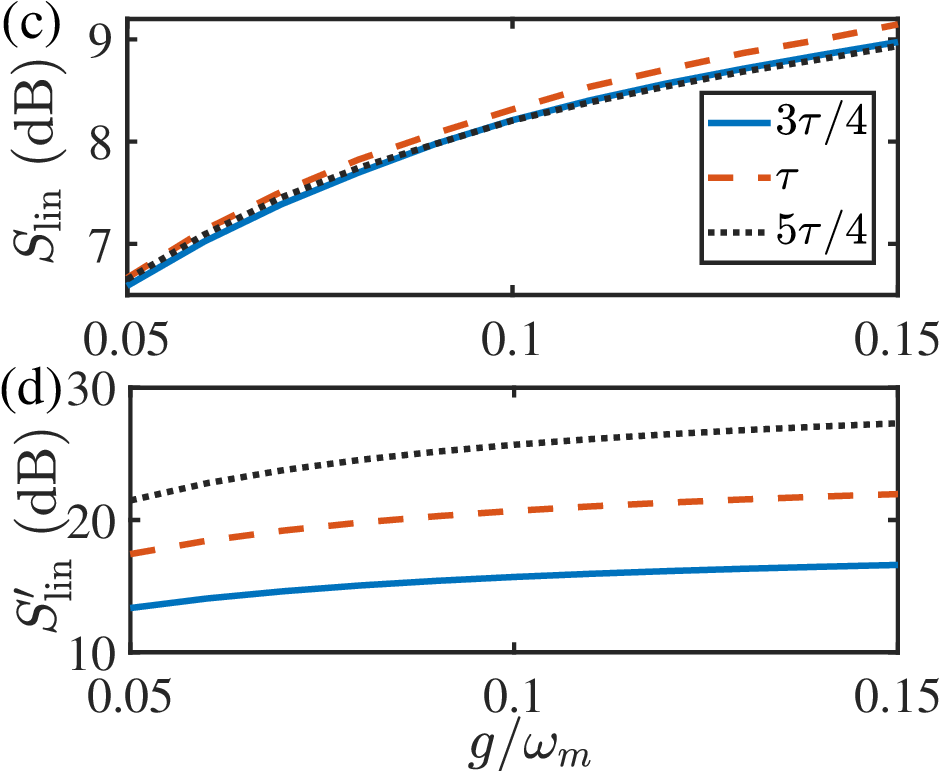}
	\caption{[(a), (b)] Time evolution of the variances $\Delta X(t)$ and $\Delta Y(t)$ using the system Hamiltonian~(\ref{Hamline}) under various coupling strengths. [(c), (d)] Squeezing level $S_{\rm lin}$ and anti-squeezing level $S'_{\rm lin}$ computed from the full system Hamiltonian~(\ref{Hamline}) at selected moments under various coupling strengths. All other parameters and initial conditions are identical to those in Fig.~\ref{vtsqueeze}.}
	\label{antisqueeze}
\end{figure}
Operating within the system instability regime results in a significant amplification of noise over time, causing the anti-squeezing effect to become increasingly pronounced during the time evolution. In this section, we analyze the anti-two-mode-squeezing effect inherent in our scheme. The anti-two-mode-squeezing operator orthogonal to the two-mode squeezing operator $X$~(\ref{quadratureX}) can be explicitly formulated as
\begin{equation}\label{quadratureY}
	Y=\cos\tilde{\phi}Y_a-\sin\tilde{\phi}X_b,
\end{equation}
which is utilized to assess the anti-two-mode-squeezing effect in our scheme. The corresponding variance of quadrature operator $Y$~(\ref{quadratureY}) can be derived as
\begin{equation}\label{Deltayphit}
	\Delta Y(t)=2C_+e^{(\Omega-\kappa_a-\kappa_b)t}+\frac{N_++1+\cos(2\tilde{\phi})N_-}{2}-2C_+,
\end{equation}
where the parameters $C_+$, $N_+$, and $N_-$ are defined in Eq.~(\ref{parameter}). When the system is unsteady, the exponential term satisfies $\Omega-\kappa_a-\kappa_b>0$, leading to an exponential divergence of the variance $\Delta Y$. Similarly, the anti-squeezing level $S'$ can be defined as $S'=10{\rm log}_{10}(\Delta Y/\Delta Y_{zp})$, where $\Delta Y_{zp}=0.5$ is the standard fluctuation in the zero-point level.

In Fig.~\ref{antisqueeze} (a) and (b), we plot the time evolution of the variances $\Delta X(t)$ and $\Delta Y(t)$, respectively. Obviously, $\Delta Y$ exhibits exponential divergence as $\Delta X$ approaches a steady value. Furthermore, as $g$ increases from $0.05\omega_m$ to $0.1\omega_m$, the variance $\Delta Y$ enhances significantly. More numerical results about the SL and anti-squeezing level are shown in Figs.~\ref{antisqueeze} (c) and (d), respectively. Consistent with the findings presented in Fig.~\ref{squeeze}, the SL $S_{\rm lin}$ does not follow a simple monotonic increase over time. Notably, the SL is maximal at time $\tau$, compared to the times $3\tau/4$ and $5\tau/4$. Conversely, the anti-squeezing level, as presented in Fig.~\ref{antisqueeze} (d), demonstrates a monotonic increase with time. Consequently, measuring the TMSS with the interval $[3\tau/4,\tau]$ yields a higher degree of squeezing while simultaneously minimizing the anti-squeezing effect. Within the optimal time window, the anti-squeezing level obtained by our scheme is comparable to that obtained in previous reservoir-engineering approaches under system stability conditions [see~\ref{appbtmsssteady}, Figure~\ref{Hamlinearize} (c)], while simultaneously enabling a significantly higher SL than those earlier methods. Moreover, according to the definition $\tau = 2\pi/(\Omega+\kappa_a+\kappa_b)$ and the expressions for $\Omega$ given in Eq.~\eqref{Omegphi}, this useful time window becomes narrower as the coupling strength increases.

Furthermore, the enhancement of anti-squeezing imposes a stricter requirement on the angular control of the measurement operator. Specifically, consider the actual measurement operator $O=\cos\theta X+\sin\theta Y$, which coincides with the ideal squeezing operator $X$ at $\theta=0$. For a misalignment $\theta\neq 0$, the variance becomes $\Delta O=\cos^2\theta\Delta X+\sin^2\theta\Delta Y$. Therefore, a larger anti-squeezing variance $\Delta Y$ implies a smaller allowable misalignment angle $\theta$ for maintaining $\Delta O<1/2$, leading to more stringent requirements on angle control. This further highlights the importance of selecting an appropriate time window in which the anti-squeezing remains within an acceptable range.

\subsection{Robustness to systematic errors}\label{secsyserror}
In the ideal situation, our TPS protocol assisted by the mechanical mode relies on the precise realization of the effective Hamiltonian given in Eq.~(\ref{Heff}). However, in practice, experimental imperfections and technical limitations prevent perfect control over the parameters of the full Hamiltonian~(\ref{Hamline}). In this section, we analyze the systematic errors arising from the driving-enhanced optomechanical coupling strengths and the mismatch of the photon frequency detunings. Specifically, the systematic error in the optomechanical coupling strengths originates from the fluctuations in the Rabi frequency $\Omega_o,o=a,b$ of the driving $H_d$ in Eq.~(\ref{Hamori}), while the detuning deviation is primarily due to the inaccuracies in the laser driving frequency $\epsilon_o$. The Hamiltonian implemented in experiments is assumed to be
\begin{equation}\label{Hamexpsys}
	H_{\rm exp}=H_{\rm int}+\gamma H_g+\eta H_{\Delta}
\end{equation}
where $H_{\rm int}$ is the ideal full system Hamiltonian in Eq.~(\ref{Hamline}) and $H_g$ and $H_{\Delta}$ are,
\begin{equation}
	\begin{aligned}
		H_g&=g(e^{-i\theta_a}a+e^{i\theta_a}a^\dag)(m+m^\dag)-G(e^{-i\theta_b}b+e^{i\theta_b}b^\dag)(m+m^\dag),\\
		H_\Delta&=\Delta_a a^\dag a-\Delta_b b^\dag b,
	\end{aligned}
\end{equation}
where $\gamma$ and $\eta$ are dimensionless perturbation coefficients for the coupling strength and the frequency detuning, respectively.
\begin{figure}[t]
\centering
\includegraphics[width=0.235\textwidth]{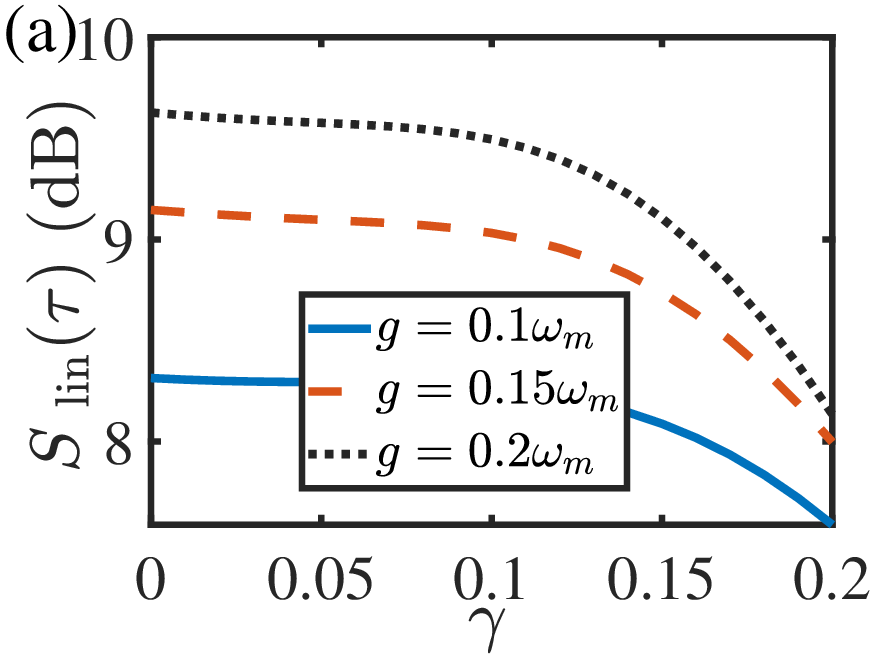}
\includegraphics[width=0.235\textwidth]{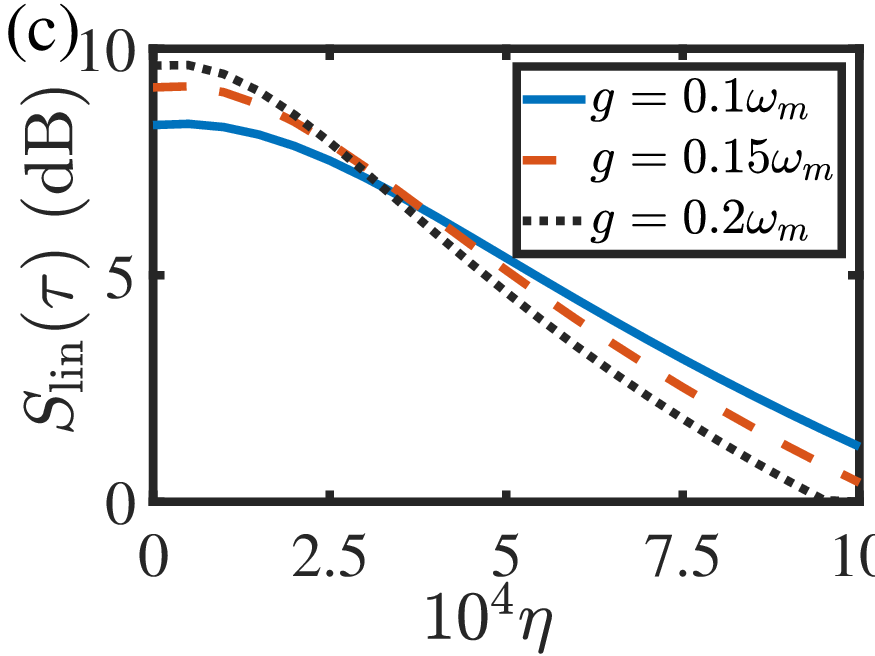}
\includegraphics[width=0.235\textwidth]{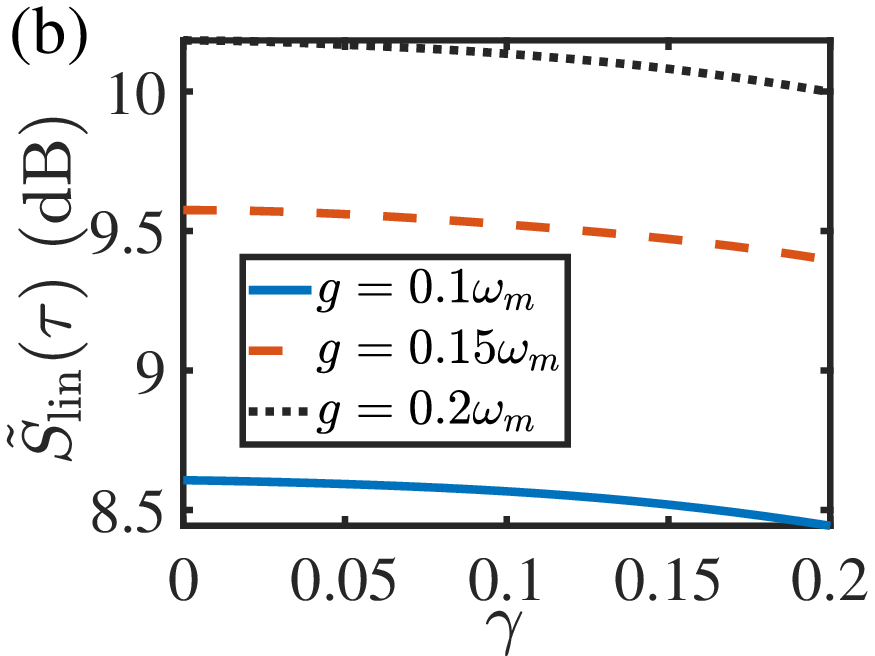}
\includegraphics[width=0.235\textwidth]{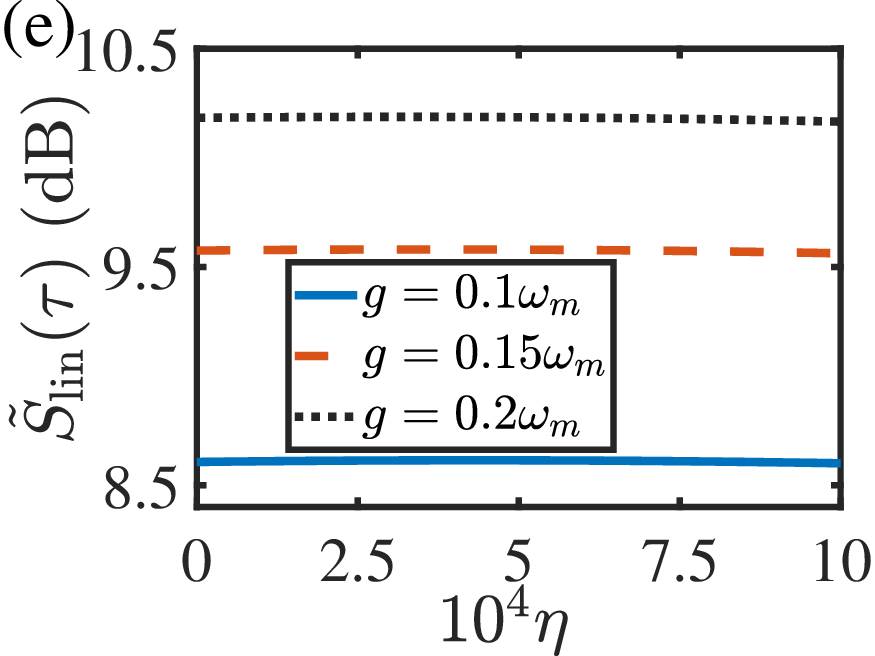}
\includegraphics[width=0.235\textwidth]{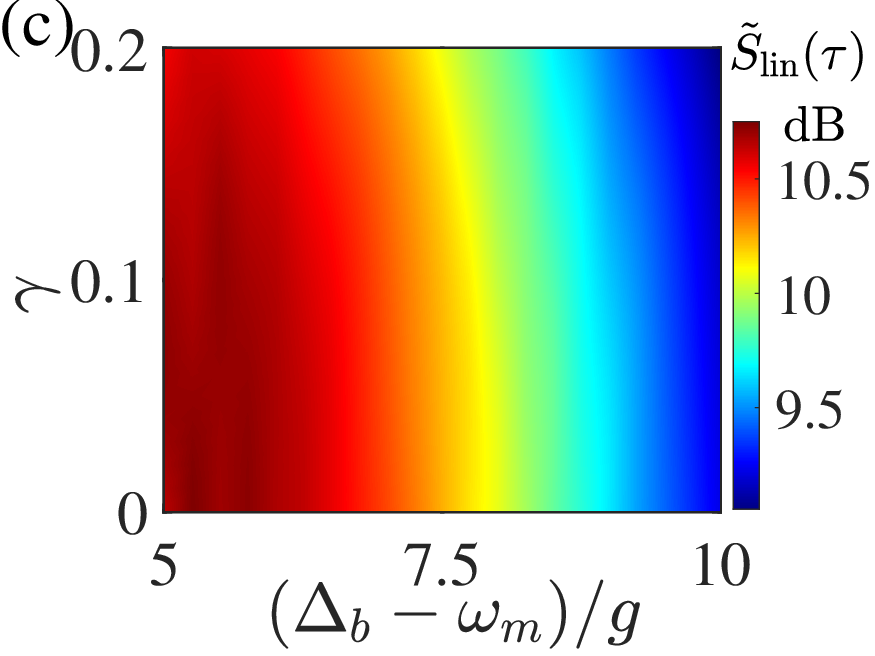}
\includegraphics[width=0.235\textwidth]{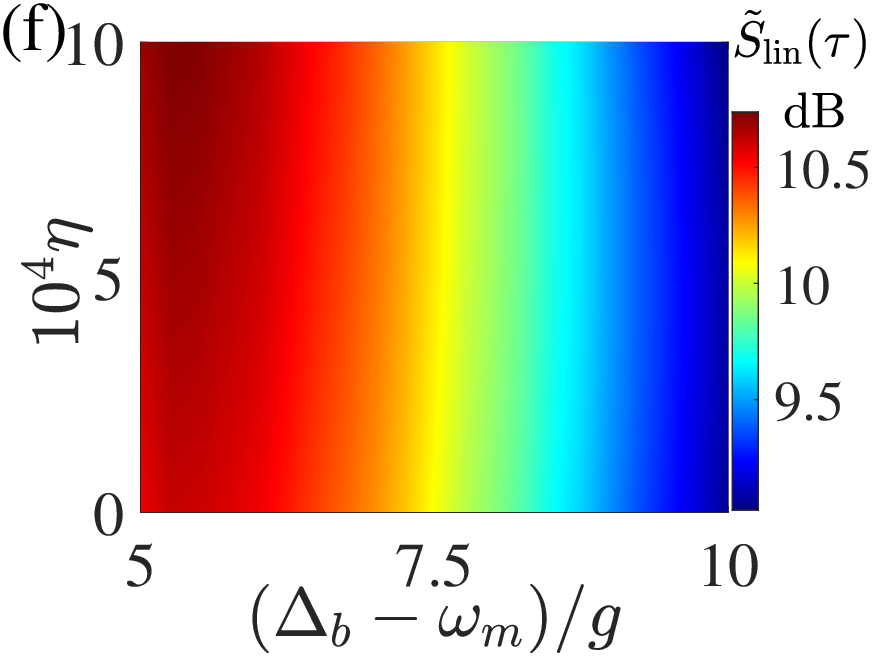}
\caption{[(a),(b)] The SL $S_{\rm lin}$ and $\tilde{S}_{\rm lin}$ at time $\tau$ evaluated under the system Hamiltonian~\eqref{Hamexpsys} as a function of the systematic error associated with the optomechanical coupling strength, respectively. (c) The numerical optical SL $\tilde{S}_{\rm lin}(\tau)$ under the space spanned by the detuning $\Delta_b$ and coupling error magnitude $\gamma$. [(d), (e)] The SL $S_{\rm lin}(\tau)$ and $\tilde{S}_{\rm lin}(\tau)$ under the system Hamiltonian~\eqref{Hamexpsys} as a function of the systematic error associated with the detuning frequency, respectively. (f) The SL $\tilde{S}_{\rm lin}(\tau)$ under the space spanned by the detuning $\Delta_b$ and detuning error magnitude $\eta$. For panels (a),(b),(d), and (e), the detuning is fixed at $\Delta_b=\omega_m+10g$. For panels (c) and (f), the coupling strengths are set as $g=G=0.1\omega_m$. All other parameters and initial system states are identical to those in Fig.~\ref{vtsqueeze}.}
\label{syserror}
\end{figure}

Under the Hamiltonian in Eq.~(\ref{Hamexpsys}) with systematic errors, the SL $S_{\rm lin}$ corresponding to the quadrature operator $X$~(\ref{quadratureX}), as well as the numerically obtained optimal $\tilde{S}_{\rm lin}$, are presented in Fig.~\ref{syserror}. It is found in Fig.~\ref{syserror} (a) that the SL $S_{\rm lin}$ sensitivity to the coupling strength error $\gamma$ is amplified with increasing strength $g$. However, when $\gamma\le 0.1$, all cases exhibit strong robustness against such errors. In contrast, as shown in Fig.~\ref{syserror} (b), the optimized SL $\tilde{S}_{\rm lin}$ are roughly invariant in the presence of errors $\gamma$, even when the error magnitude reaches $20\%$. Fig.~\ref{syserror} (c) presents the SL $\tilde{S}_{\rm lin}$ under various detuning $\Delta_b$ and coupling error magnitude $\gamma$. It can be observed that smaller detuning lead to higher SL. This behavior can be understood as a consequence of smaller detuning resulting in larger effective coupling $g_{\rm eff}$~(\ref{geffdelta}), regardless of the error magnitude. Therefore, we conclude that systematic errors in the optomechanical coupling have a slight influence on the selection of the optimal quadrature operator, causing a slight deviation from the theoretical prediction $X$~(\ref{quadratureX}) while leaving the two-mode squeezing phenomenon largely unaffected.

Similarly, the SL robustness against the frequency deviation of the two photonic modes is presented in Figs.~\ref{syserror} (d), (e), and (f). As shown in Fig.~\ref{syserror} (d), the SL $S_{\rm lin}$ decreases significantly with increasing error magnitude $\eta$, and the sensitivity becomes more pronounced for larger values of $g$. For instance, when $g=0.2\omega_m$, the $S_{\rm lin}$ declines into $0$ as $\eta$ approaches $10^{-3}$. This indicates that detuning errors have a substantial impact on the achievable $S_{\rm lin}$, which is more detrimental than errors in the coupling strength. However, as seen in Figs.~\ref{syserror} (e) and (f), this deviation has no significant effect on the optimized SL $\tilde{S}_{\rm lin}$. Theoretically, the quadrature operator $X$~(\ref{quadratureX}), derived using the effective Hamiltonian, is obtained under a strict condition, i.e., $\Delta_a=-\Delta_b+\delta$, where $\delta\sim0.01-0.1\Delta_b$ (see~\ref{appcHeff}). As a result, frequency deviations between the two photonic modes can significantly affect the selection of the optimal quadrature operator. Nevertheless, such deviations do not prevent the emergence of a stable TPS, as the effective TPS coupling $g_{\rm eff}$ can still be successfully constructed when $\Delta_a\approx-\Delta_b$. In fact, under this condition, the system is governed by the effective Hamiltonian given in Eq.~(\ref{Heffinteract}), rather than the $H_{\rm eff}$ shown in Eq.~(\ref{Heff}). 

\subsection{Feasibility in thermal environment noises}\label{secthermal}
\begin{figure}[ht]
	\centering
	\includegraphics[width=0.235\textwidth]{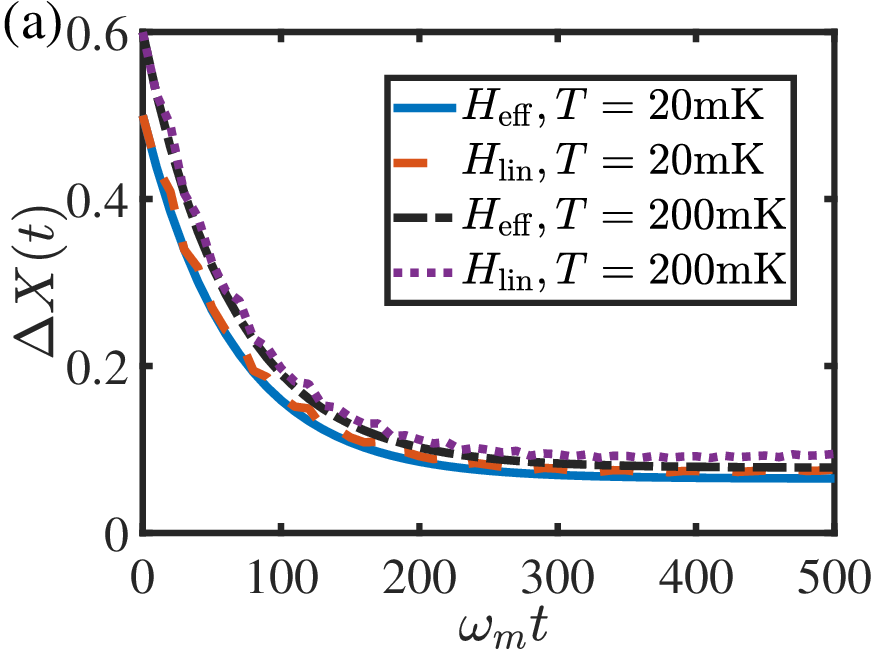}
	\includegraphics[width=0.235\textwidth]{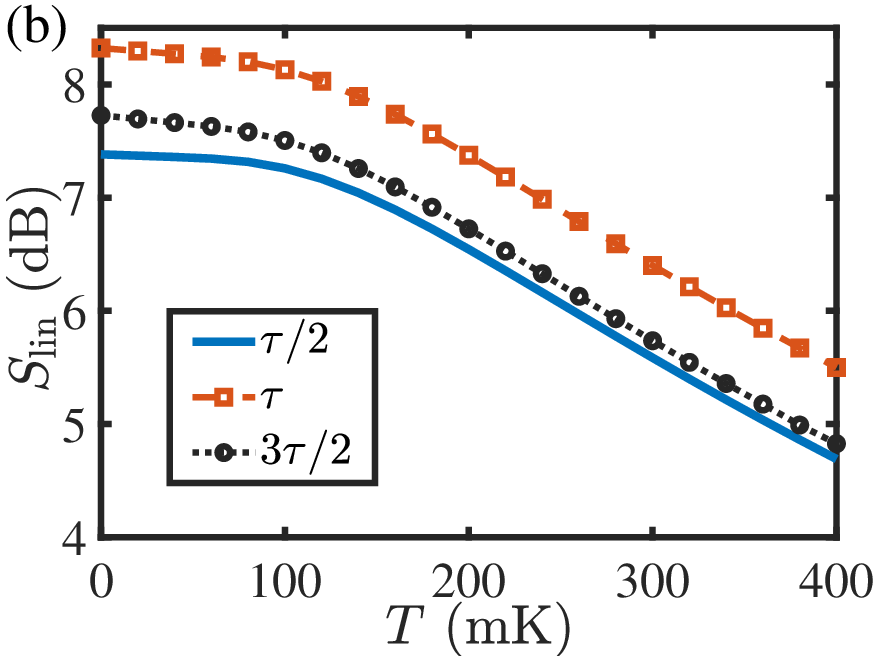}
	\caption{(a) Time evolution of $\Delta X(t)$ obtained using either the effective Hamiltonian~(\ref{Heff}) or the full system Hamiltonian~(\ref{Hamline}) at various temperatures. (b) The SL computed from the full system Hamiltonian~(\ref{Hamline}) at selected moments under different temperatures. All other parameters are identical to those in Fig.~\ref{vtsqueeze}, except for the thermal occupation numbers. Specifically, the photon modes are initially prepared in thermal equilibrium states, while the phonon mode $m$ is initially in the vacuum state.}
	\label{thermal}
\end{figure}
Our protocol is primarily focused on generating phonon-assisted two-photon-mode squeezing in the specific cavity optomechanical system. In the preceding numerical simulations, we assume the thermal occupations of photon and phonon modes to be $N_a=N_b=0$ and $N_m=10$, respectively, approximate values corresponding to a low temperature of $T\sim 10$ mK. In this section, we analyze the performance of two-mode squeezing across a range of temperatures, employing a representative set of experimentally accessible parameters. Specifically, we consider microwave resonator transition frequencies of $\omega_a/2\pi=\omega_b/2\pi=10$ GHz and a mechanical mode frequency of $\omega_m/2\pi=10$ MHz in our simulations. The corresponding numerical results are presented in Fig.~\ref{thermal}.

Figure~\ref{thermal} (a) compares the variance dynamics $\Delta X(t)$ with the effective Hamiltonian~(\ref{Heff}) and the full system Hamiltonian~(\ref{Hamline}) under varying thermal conditions. Within the time interval, $\omega_mt\le 500$, the analytical results from Eq.~(\ref{Deltaxphit}) show excellent agreement with the corresponding numerical results derived from Eq.~(\ref{wholedotV}). As the temperature $T$ increases from $20$ mK to $200$ mK, $\Delta X(t)$ shows only a slight increase, indicating that the high-quality TPS is well preserved even at elevated temperatures. In Fig.~\ref{thermal} (b), we plot the SL $S_{\rm lin}(\tau/2)$, $S_{\rm lin}(\tau)$, and $S_{\rm lin}(3\tau/2)$ as a function of temperature, using the full system Hamiltonian. The results demonstrate that the protocol sustains a high SL, with $S_{\rm lin}(\tau)\ge 8$ dB, at low temperatures $T\le 100$ mK. As the temperature gradually increases, the SL $S_{\rm lin}(\tau)$ gradually declines. However, a substantial squeezing of $5.5$ dB is still achieved at $T=100$ mK. Besides, consistent with the results in Fig.~\ref{squeeze}, both $S_{\rm lin}(\tau/2)$ and $S_{\rm lin}(3\tau/2)$ are approximately $1$ dB lower than $S_{\rm lin}(\tau)$ across the considered temperature range. 

\section{Discussion and Conclusion}\label{Secexpercon}
Our protocol is designed for hybrid cavity optomechanical systems. In recent experiments~\cite{cohermechanical,entmechanical,squeemech,Superposition,coherentcoupling,active,map}, the phonon frequency $\omega_m/2\pi\sim 10-100$ MHz, with a low decay rate $\kappa_m/2\pi\sim 10-100$ Hz, corresponding to $\kappa_m/\omega_m\sim 10^{-6}$. The cavity quality factor $Q=\omega/\kappa$, where $\omega$ is the transition frequency and $\kappa$ is the decay rate, depends on the specific cavity structure. Fabry-P\'erot~\cite{Rempe:92} and whispering-gallery-mode~\cite{Gorodetsky:96} (micro) cavities exhibit ultrahigh quality factors on the order of $10^{10}$, with decay rates $\kappa/2\pi \sim 10$ KHz. For microwave resonators, the transition frequency of the photon mode is typically around $10$ GHz, with a high-quality factor $Q\sim 10^{4}- 10^{7}$. The corresponding photon loss rates are $\kappa/2\pi\sim 10^{-3}\omega_m$. The Rabi frequency of external driving is defined as $\Omega\equiv \sqrt{\kappa P_d/(\hbar\omega_d)}$~\cite{synergizing}, where $P_d$ and $\omega_d$ represent the power and frequency of the drive, respectively. In recent experiments~\cite{MechanicalBis}, $\kappa/2\pi\sim 0.01-0.1$ MHz and $P_d\sim 20-30$ dBm ($100-1000$ mW), yielding a Rabi frequency $\Omega\sim 10^{14}-10^{15}$ Hz, and consequently, the mean excitation value shown in Eq.~(\ref{steadyope}) is about $|\alpha|\approx 10^{7}\sim 10^{8}$. The driving-enhanced photon-phonon coupling is given by $g\equiv g_a\alpha \sim 0.1\omega_m$, where the single excitation coupling $g_a$ is about $1-10$ Hz. As shown in Sec.~\ref {secsyserror}, the coupling strength $g$ can tolerate systematic errors of up to $\sim 10\%$ ($\sim 0.1-1$ MHz). Owing to the dependence of $g$ on the driving power $P_d$, the scheme is robust against power fluctuations at the $1\%$ level. Moreover, it remains valid for driving-frequency deviations of $10-100$ kHz. The thermal occupations of photon modes and phonons are respectively $N_a\approx N_b\approx 0$ and $N_m\approx 10$ at $T\sim 10$ mK. All of the values are consistent with the parameters used in Secs.~\ref {Sectms},~\ref{secanti}, and~\ref{secsyserror}. In addition, beyond the cavity optomechanical system, this scheme is extendable to other hybrid platforms involving three bosonic modes. For example, a magnonic interface could be employed to realize microwave–optical squeezing~\cite{Nonrecisteer}.

In summary, we have proposed a protocol for generating TMSS in a three-mode cavity optomechanical system, wherein a mechanical resonator is simultaneously coupled to two microwave (or optical) modes. The protocol relies on an effective two-mode squeezing Hamiltonian mediated by the mechanical mode. To numerically validate the effective Hamiltonian in the presence of nonconservative excitations, we employ a method based on diagonalizing the full system’s transition matrix. Within the open quantum systems framework, we derive the dynamical evolution of TMSS generation governed by the effective Hamiltonian. Our analysis demonstrates that stable TMSS with a high SL can be achieved even beyond the system stability regime, with high-SL TMSS dynamically generated within an appropriate time window while maintaining the anti-squeezing at a controlled level. Moreover, the protocol exhibits robustness against systematic errors and thermal environmental noises. It offers notable advantages in terms of system controllability and provides a practical route to implementing TMSS generation in the presence of environmental noises and technical imperfections. Overall, our protocol establishes a versatile and scalable framework for TMSS generation, with broad potential applications in quantum information processing and quantum precision measurement using bosonic systems.

\section{Data availability statement}
Data will be made available on request.

\section{Acknowledgements} 
We acknowledge financial support from the National Science Foundation of China (12404405) and the Science Foundation of Hebei Normal University of China (L2024B10).

\appendix
\section{System linearized Hamiltonian}\label{appaline}
This appendix contributes to deriving the linearized Hamiltonian in Eq.~(\ref{Hamline}). With respect to the transformation $U(t)=\exp{(i\epsilon_ata^\dag a+i\epsilon_btb^\dag b)}$, the original Hamiltonian in Eq.~(\ref{Hamori}) turns out to be 
\begin{equation}\label{Ham1}
H_s=\omega_mm^\dag m+\sum_{o=a,b}\Delta_oo^\dag o+g_oo^\dag o(m+m^\dag)+\Omega_o(o+o^\dag),
\end{equation}
where $\Delta_o=\omega_o-\epsilon_o$. Using the quantum Langevin equation, the time evolution of the system operators can be written as
\begin{equation}\label{qleoperator}
\begin{aligned}
	\dot{a}&=-(i\Delta_a+\kappa_a)a-ig_aa(m+m^\dag)-i\Omega_a+\sqrt{2\kappa_a}a_{in},\\
	\dot{b}&=-(i\Delta_b+\kappa_b)b-ig_bb(m+m^\dag)-i\Omega_b+\sqrt{2\kappa_b}b_{in},\\
	\dot{m}&=-(i\omega_m+\kappa_m)m-ig_aa^\dag a-ig_bb^\dag b+\sqrt{2\kappa_m}m_{in},\\
\end{aligned}
\end{equation}
where $\kappa_a$, $\kappa_b$, and $\kappa_m$ are the decay rates of the modes $a$, $b$, and $m$, respectively. $o_{in}, o=a,b,m$ is the input noise operator for the mode $o$, which is characterized by the covariance functions: $\langle o_{in}(t)o^\dag_{in}(t')\rangle=[N_o+1]\delta(t-t')$ and $\langle o^\dag_{in}(t)o_{in}(t')\rangle=N_o\delta(t-t')$, under the Markovian approximation. $N_o=[\exp(\hbar\omega_o/k_BT)-1]^{-1}$ is the thermal occupation number of mode $o$.

Under proper strong driving pulses, by performing the standard linearization approach, we write the operators as their classical values plus small fluctuations, i.e., $a=\alpha+\delta a$, $b=\beta+\delta b$, and $m=M+\delta m$. Here, $\alpha,\beta, M$ are the complex numbers and $\delta o, o=a,b,m$ are the fluctuation operators. The classical values are determined by
\begin{equation}\label{steadyoperator}
\begin{aligned}
	\dot{\alpha}&=-(i\Delta_a+\kappa_a)\alpha-ig_a\alpha(M+M^*)-i\Omega_a,\\
	\dot{\beta}&=-(i\Delta_b+\kappa_b)\beta-ig_b\beta(M+M^*)-i\Omega_b,\\
	\dot{M}&=-(i\omega_m+\kappa_m)M-ig_a|\alpha|^2-ig_b|\beta|^2,\\
\end{aligned}
\end{equation}
It is important to note that $\alpha, \beta, M$ can, in principle, achieve any desirable values by time-dependent modulation of the corresponding driving fields $\Omega_a$ and $\Omega_b$. These quantities rapidly converge to their respective constant magnitudes through appropriate tuning of the Rabi frequencies, which occurs on a timescale much faster than that of the fluctuation dynamics. The classical constants are obtained by setting the time derivatives in Eq.~(\ref{steadyoperator}) to zero, which satisfy
\begin{equation}\label{steadyope}
\begin{aligned}
\alpha&=-\frac{\Omega_a}{\Delta_a-i\kappa_a-2g_a{\rm Re}(M)},\\
\beta&=-\frac{\Omega_b}{\Delta_b-i\kappa_b-2g_b{\rm Re}(M)},\quad M=-\frac{g_a|\alpha|^2+g_b|\beta|^2}{\omega_m-i\kappa_m},
\end{aligned}
\end{equation}
where ${\rm Re}(M)$ is the real part of $M$. When $\kappa_a,\kappa_b,\kappa_m\ll|\Delta_a|,|\Delta_b|,\omega_m$ and both $g_a$ and $g_b$ are significantly small, the classical constants of photon modes $a$ and $b$ approximately equal to $|\alpha|\approx\Omega_a/\Delta_a$ and $|\beta|\approx \Omega_b/\Delta_b$, respectively.

By substituting these classical constants in Eq.~(\ref{steadyope}) into the Eq.~(\ref{qleoperator}) and ignoring all the high-order terms of fluctuations, the quantum Langevin equations describing the fluctuation operator $\delta o$ can be written as
\begin{equation}\label{flucope}
\begin{aligned}
\dot{\delta a}&=-[i\Delta_a+2ig_a{\rm Re}(M)+\kappa_a]\delta a-ig_a\alpha(\delta m+\delta m^\dag)+\sqrt{2\kappa_a}a_{in},\\
\dot{\delta b}&=-[i\Delta_b+2ig_b{\rm Re}(M)+\kappa_b]\delta b-ig_b\beta(\delta m+\delta m^\dag)+\sqrt{2\kappa_b}b_{in},\\
\dot{\delta m}&=-(i\omega_m+\kappa_m)\delta m-ig_a(\alpha a^\dag+\alpha^*a)-ig_b(\beta b^\dag+\beta^*b)\\
&+\sqrt{2\kappa_m}m_{in}.\\
\end{aligned}
\end{equation}
The corresponding effective linearized Hamiltonian can be described as
\begin{equation}\label{hamlineapp}
\begin{aligned}
	H_{\rm lin}&=\tilde{\Delta}_a\delta a^\dag\delta a+\tilde{\Delta}_b\delta b^\dag\delta b+\omega_m\delta m^\dag\delta m\\
	&+(g^*\delta a+g\delta a^\dag)(m+m^\dag)+(G^*\delta b+G\delta b^\dag)(m+m^\dag),
\end{aligned}
\end{equation}
where $g=g_a\alpha\approx g_a\Omega_a/\Delta_a$ and $G=g_b\beta\approx g_b\Omega_b/\Delta_b$ are the driving-enhanced coupling strengths. The modified detunings $\tilde{\Delta}_a=\Delta_a-2g_a{\rm Re}(M)\approx\Delta_a$ and $\tilde{\Delta}_b=\Delta_b-2g_b{\rm Re}(M)\approx\Delta_b$ provided by $g_o{\rm Re}(M)\ll\Delta_o, o=a,b$. It is the linearized Hamiltonian in Eq.~(\ref{Hamline}) in the main text. For simplicity and with no loss of generality, we apply the convention $\tilde{\Delta}_o\to\Delta_o$, $\delta o\to o, o=a,m,b$, and $g\to ge^{-i\theta_a}$, $G\to Ge^{-i\theta_b}$ in the main manuscript and following content.

\section{Two-mode squeezing via reservoir engineering method}\label{appbtmsssteady}
Under the parameter conditions $\Delta_a=\omega_m, \Delta_b=-\omega_m,$ and $g, G\ll \Delta_a,\Delta_b,\omega_m$, after the rotating wave approximation, the linearized system Hamiltonian $H_{\rm lin}$ turns into
\begin{equation}\label{Hintappb}
	H_{\rm int}=g(e^{i\theta_a}a^\dag m+e^{-i\theta_a}am^\dag)+G(e^{-i\theta_b}bm+e^{i\theta_b}b^\dag m^\dag).
\end{equation} 
In the regime $g>G$, one can introduce a Bogoliubov mode $\tilde{a}$ to transform the above Hamiltonian, defined as
\begin{equation}
	\tilde{a}=e^{-i\theta_a}a\cosh r+e^{i\theta_b}b^\dag\sinh r=S(r)aS^\dag (r),
\end{equation}
where $S(r)\equiv\exp(r^*ab-ra^\dag b^\dag)$, $r=|r|e^{i(\theta_a+\theta_b)}$, and $\tanh |r|=G/g$. The joint vacuum of $\tilde{a}$ is the TMSS $|\tilde{0}\rangle=S(r)|00\rangle$, where $|00\rangle\equiv|0\rangle_a|0\rangle_b$. Consequently, reservoir engineering that cools $\tilde{a}$ into its vacuum state deterministically generates a TMSS of the original modes $a$ and $b$.

Using the quantum Langevin equation, the dynamics of the quantum system under the interaction Hamiltonian~(\ref{Hintappb}) can be written in a matrix form
\begin{equation}\label{uat}
	\dot{u}(t)=A_{\rm int}u(t)+n(t),
\end{equation}
where $u^T(t)=[X_a(t),Y_a(t),X_b(t),Y_b(t),X_m(t),Y_m(t)]$ is the vector of quadrature fluctuation operators, and $X_o=(e^{-i\theta_o}o+e^{i\theta_o}o^\dag)/\sqrt{2}$, $Y_o=(e^{-i\theta_o}o-e^{i\theta_o}o^\dag)/i\sqrt{2}$, $o=a,m,b$, $\theta_m=0$. $n^T(t)=[X^{in}_a(t),Y^{in}_a(t),X^{in}_b(t),Y^{in}_b(t),X^{in}_m(t),Y^{in}_m(t)]$ is the vector of corresponding noise operators, and $X^{in}_o=(e^{-i\theta_o}o_{in}+e^{i\theta_o}o_{in}^\dag)/\sqrt{2}$, $Y^{in}_o=(e^{-i\theta_o}o_{in}-e^{i\theta_o}o_{in}^\dag)/i\sqrt{2}$. The transition matrix $A_{\rm int}$ is
\begin{equation}\label{alinmatrix}
\begin{aligned}
	A_{\rm int}=\begin{bmatrix}
		-\kappa_a & 0 & 0 & 0 &0 & g\\
		0 &-\kappa_a &0 & 0 &-g & 0\\
		0 &0 &-\kappa_b &0 &0 & -G\\
		0 &0 &0 &-\kappa_b &-G & 0\\
		0 &g &0 &-G &-\kappa_m & 0\\
		-g &0 &-G &0 &0& -\kappa_m\\
	\end{bmatrix}
\end{aligned}
\end{equation}

Due to the above dynamics in Eq.~(\ref{uat}) and the zero-mean Gaussian nature of the quantum noises, the hybrid system evolves as a Gaussian state, which can be completely characterized by a $6\times 6$ CM $V(t)$. The dynamics of CM $V(t)$ satisfies 
\begin{equation}
	\dot{V}(t)=A_{\rm int}V(t)+V(t)A_{\rm int}^T+D.
\end{equation}
The element of $V(t)$ is given by $V_{ij}=\langle u_i(t)u_j(t)+u_j(t)u_i(t)\rangle/2$, where $u_i(t)$ is the $i$ term of $u(t)$ and $i=1,2,\cdots,6$. $D$ is the noise covariance matrix, where the diagonal elements are 
$D_{11}=D_{22}=\kappa_a(2N_a+1)$, $D_{33}=D_{44}=\kappa_b(2N_b+1)$, and $D_{55}=D_{66}=\kappa_m(2N_m+1)$, all non-diagonal elements are zero.

The steady-state CM can be achieved by setting $\dot{V}(t)=0$ and its elements are
\begin{equation}\label{cmsteadyele}
\begin{aligned}
V_{11}&=\frac{1}{2}+\frac{G^2g^2(\kappa_m+2\kappa_a)}{\tilde{v}},\quad\\
V_{33}&=\frac{1}{2}+\frac{G^2[2(\kappa_a+\kappa_m)(g^2+\kappa_a\kappa_m)-G^2\kappa_m)]}{\tilde{v}},\\
V_{66}&=\frac{1}{2}+\frac{G^2\kappa_a[g^2-G^2+2\kappa_a(\kappa_a+\kappa_m)]}{\tilde{v}},\quad\\
V_{13}&=-\frac{Gg[G^2\kappa_a+(g^2+\kappa_a\kappa_m)(\kappa_a+\kappa_m)]}{\tilde{v}},\\
V_{16}&=-\frac{gG^2\kappa_a(2\kappa_a+\kappa_m)}{\tilde{v}},\quad\\
V_{36}&=-\frac{G[2\kappa_a(\kappa_a+\kappa_m)(g^2+\kappa_a\kappa_m)-G^2\kappa_a\kappa_m]}{\tilde{v}},\\\end{aligned}
\end{equation}
where 
\begin{equation}
	\tilde{v}=(\kappa_a+\kappa_m)(G^2-g^2-\kappa_a\kappa_m)[G^2-g^2-2\kappa_a(\kappa_a+\kappa_m)],
\end{equation}
and non-zero matrix elements are $V_{22}=V_{11}$, $V_{44}=V_{33}$, $V_{31}=V_{13}$, $V_{24}=V_{42}=-V_{13}$, $V_{61}=V_{16}$, $V_{25}=V_{52}=-V_{26}$, and $V_{45}=V_{54}=V_{63}=V_{36}$. For simplicity and with no loss of generality, here we assume $\kappa_b=\kappa_a$ and $N_a=N_b=N_m=0$. 

\begin{figure}[b]
	\centering
	\includegraphics[width=0.235\textwidth]{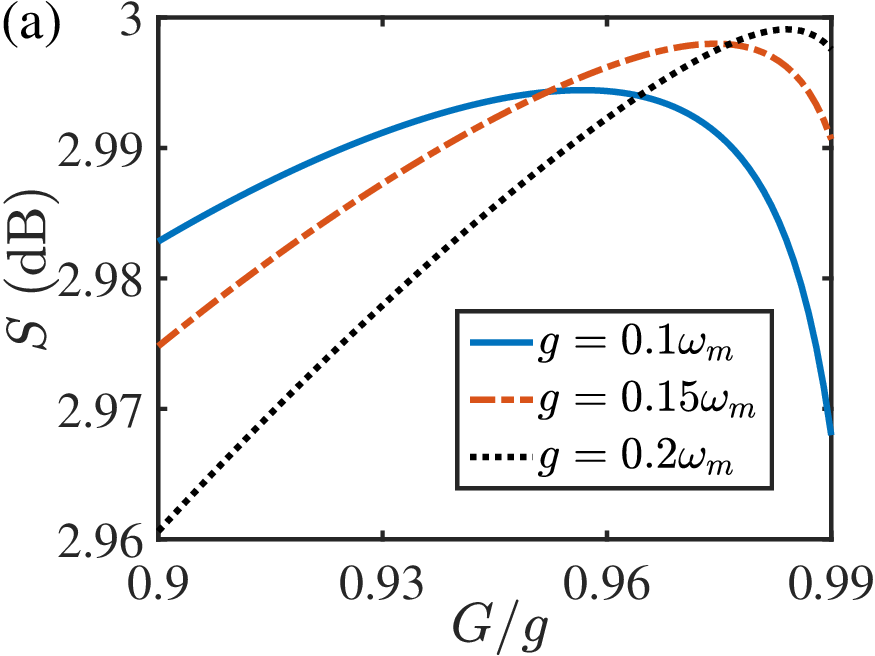}
	\includegraphics[width=0.235\textwidth]{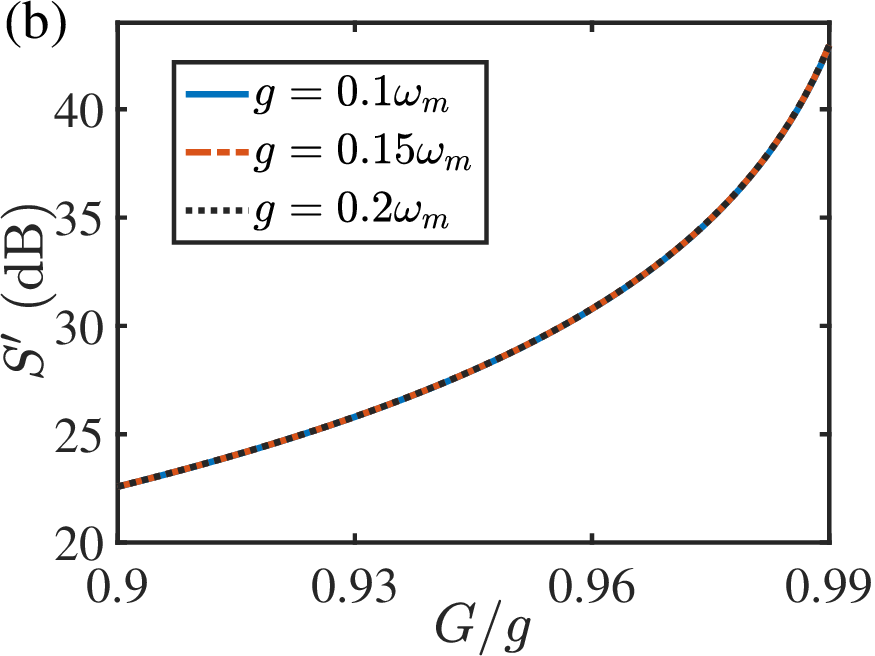}
	\caption{[(a), (b)] The SL $S$ and anti-squeezing level $S'$ as a function of the coupling strength $G/g$ using the whole system Hamiltonian in Eq.~(\ref{Hamline}), respectively. Here, the parameters are set as $\Delta_a=\omega_m$, $\Delta_b=-\omega_m$, $\kappa_a=\kappa_b=10^{-3}\omega_m$, $\kappa_m=10^{-6}\omega_m$, and the thermal occupation numbers $N_a=N_b$ and $N_m=10$.}\label{Hamlinearize}
\end{figure}

The two-mode squeezing operator can be written as $X=(X_a+X_b)/\sqrt{2}$. Then, with the CM elements shown in Eq.~(\ref{cmsteadyele}), its variance 
$\Delta X=\langle X\rangle^2-\langle X\rangle^2$ can be described as
\begin{equation}
\begin{aligned}
\Delta X&=\frac{1}{2}+\frac{G^2(g^2-G^2)\kappa_m}{2\tilde{v}}+\frac{G^2(g^2-Gg)\kappa_a}{\tilde{v}}\\
&+\frac{(G^2-Gg)(g^2+\kappa_a\kappa_m)(\kappa_a+\kappa_m)}{\tilde{v}}.
\end{aligned}
\end{equation}
In the strong coupling regime and the decay rate of photon is larger than the one of phonon~\cite{cohermechanical,entmechanical,squeemech,Superposition,coherentcoupling,active,map}, i.e., $g, G>\kappa_a\gg\kappa_m$, the variance $\Delta X$ can be approximated as
\begin{equation}\label{Deltaxapprox}
\Delta X\approx \frac{1}{2}-\frac{Gg}{(G+g)^2}.
\end{equation}
The variance reaches its minimal value, $\Delta X_{\rm min}=1/4$, when $G\to g$, which corresponds to a SL of $3$ dB. The SL $S$ in the decibel unit is defined by $S=-10\log_{10}(\Delta X/\Delta X_{zp})$, where $\Delta X_{zp}=0.5$ is the standard fluctuation in the zero-point level. Obviously, a smaller $\Delta X$ yields a higher SL $S$. In practical implementations, cooling the phonon mode to its ground state remains a challenging task. Moreover, a higher phonon occupation number $N_m$ degrades the squeezing performance. Consequently, under realistic conditions, the achievable SL of this method is difficult to exceed the $3$ dB limit.

In Figs.~\ref{Hamlinearize}(a) and (b), we present the numerical results for the SL $S$ and the anti-squeezing level $S'$ under different coupling strengths, respectively. The anti-squeezing level is defined as $S'=10\log_{10}(\Delta Y/\Delta Y_{zp})$, where $\Delta Y$ is the variance of the anti-squeezing operator $Y=(Y_a+Y_b)/\sqrt{2}$ and $\Delta Y_{zp}=0.5$ denotes the zero-point fluctuation. As shown in Fig.~\ref{Hamlinearize}(a), the SL $S$ remains below the $3$ dB limit for the chosen parameters. With increasing coupling strength $g$, the optimal ratio $G/g$ (corresponding to the inflection point where maximal squeezing occurs) shifts to larger values, accompanied by an enhanced achievable SL. However, for values of $G/g$ far from this optimum, further increasing $g$ instead suppresses the squeezing. Figure~\ref{Hamlinearize}(b) shows the corresponding anti-squeezing level. The three curves nearly overlap and lie within the range of $22-45$ dB, indicating that the anti-squeezing is largely insensitive to variations in $g$. In addition, the anti-squeezing level increases monotonically with $G/g$.

\section{The effective Hamiltonian}\label{appcHeff}
This appendix contributes to deriving the effective Hamiltonian in Eq.~(\ref{Heff}) and evaluates its validity. When the detuning frequency of mode $a$ is almost opposite the detuning of mode $b$, and both of them are far resonant from the frequency of phonon, i.e., $\Delta_a+\Delta_b\approx 0$ and $|\Delta_a-\omega_m|, |\Delta_b-\omega_m|\gg g, G$, the interaction Hamiltonian $V$ in Eq.~\eqref{Hamline} can be regarded as a perturbation to the free Hamiltonian $H_0$. Under these conditions, the effective TPS coupling between photons $a$ and $b$, mediated by phonon mode $m$, can be successfully constructed. The detailed derivations are provided below.

With respect to $U(t)=\exp(-iH_0t)$, the linearized Hamiltonian $H_{\rm lin}$ in Eq.~\eqref{Hamline} can be transformed to the interaction picture
\begin{equation}\label{Hamint}
\begin{aligned}
	H_I&=ge^{-i\theta_a}ame^{-i(\Delta_a+\omega_m)t}+ge^{-i\theta_a}am^\dag e^{-i(\Delta_a-\omega_m)t}\\
	&\quad+Ge^{-i\theta_b}bme^{-i(\Delta_b+\omega_m)t}+Ge^{-i\theta_b}bm^\dag e^{-i(\Delta_b-\omega_m)t}+{\rm H.c.}.
\end{aligned}
\end{equation}
In the interaction picture, if the system is governed by the Hamiltonian $H_I$ of the form~\cite{james},
\begin{equation}\label{Hintform}
	H_I=\sum_s g_s[h_s\exp(i\omega_st)+h^\dag_s\exp(-i\tilde{\omega}_st)].
\end{equation}
When $|\omega_s-\tilde{\omega}_s|\ll g_s\ll|\omega_s|,|\tilde{\omega}_s|$, the effective Hamiltonian up to the second order can be derived as
\begin{equation}\label{Heffjames}
	H_{\rm eff}=\sum_s\frac{g_s^2}{2}\left(\frac{1}{\omega_s}+\frac{1}{\tilde{\omega}_s}\right)[h_s,h^\dag_s].
\end{equation}
Then, the effective Hamiltonian corresponding to the system's Hamiltonian~\eqref{Hamint} in the interaction picture can be written as 
\begin{equation}\label{Heffinteract}
\begin{aligned}
H_{\rm eff}&=\frac{2g^2\omega_m}{\Delta^2_a-\omega^2_m}a^\dag a+\frac{2G^2\omega_m}{\Delta^2_b-\omega^2_m}b^\dag b\\
&-\left(\frac{2g^2\Delta_a}{\Delta^2_a-\omega^2_m}+\frac{2G^2\Delta_b}{\Delta^2_b-\omega^2_m}\right)m^\dag m\\
&+\left(\frac{gG\omega_m}{\Delta^2_a-\omega^2_m}+\frac{gG\omega_m}{\Delta^2_b-\omega^2_m}\right)(e^{-i\theta}ab+e^{i\theta}a^\dag b^\dag),
\end{aligned}
\end{equation}
where $\theta=\theta_a+\theta_b$. Rotating it into the lab frame and discard the non-interacting phonon mode $m$, the effective Hamiltonian in Eq.~\eqref{Heffinteract} turns into
\begin{equation}\label{Heffint}
\begin{aligned}
H_{\rm eff}&=\left(\Delta_a+\frac{2g^2\omega_m}{\Delta^2_a-\omega^2_m}\right)a^\dag a+\left(\Delta_b+\frac{2G^2\omega_m}{\Delta^2_b-\omega^2_m}\right)b^\dag b\\
&\quad+\left(\frac{gG\omega_m}{\Delta^2_a-\omega^2_m}+\frac{gG\omega_m}{\Delta^2_b-\omega^2_m}\right)(e^{-i\theta}ab+e^{i\theta}a^\dag b^\dag).
\end{aligned}
\end{equation}

A purely two-mode squeezing coupling between modes $a$ and $b$ requires that the first two terms in Eq.~\eqref{Heffint} constitute an identity operator. Assuming the distance between $\Delta_a$ and $-\Delta_b$ is $\delta$, one can have 
\begin{equation}\label{delta}
	\begin{aligned}
		\delta&\equiv\Delta_a+\Delta_b=\frac{2(g^2+G^2)\omega_m}{\omega^2_m-\Delta^2_b},
	\end{aligned}
\end{equation}
up to the second order of the coupling strengths $g$ and $G$. At the condition $\Delta_a=-\Delta_b+\delta$, the effective Hamiltonian in Eq.~\eqref{Heffint} eventually turns into
\begin{equation}\label{Heffint2}
	H_{\rm eff}=\frac{2gG\omega_m}{\Delta^2_b-\omega^2_m}(e^{-i\theta}ab+e^{i\theta}a^\dag b^\dag).
\end{equation}
That is exactly the effective Hamiltonian in Eq.~\eqref{Heff} describing the coupling between photon modes $a$ and $b$. It is worth noting that choosing the effective Hamiltonian in Eq.~\eqref{Heffint2}, rather than the more general form in Eq.~\eqref{Heffint}, enables the theoretical derivation of an optimal squeezing operator in Sec.~\ref{Sectms}. The effective Hamiltonian given in Eq.~\eqref{Heffint} is already capable of generating the stable TMSS with a high SL.

Now, we evaluate the applicability range of the effective Hamiltonian in Eq.~\eqref{Heffint2} regarding the coupling strengths by diagonalizing the transition matrix of the whole system~\cite{Delocalization}. This approach differs from previous works~\cite{intermagnon,oneexcite,nonlinear} that diagonalize the full Hamiltonian in a truncated finite-dimensional Hilbert space. 

\begin{figure}[t]
	\centering
	\includegraphics[width=0.235\textwidth]{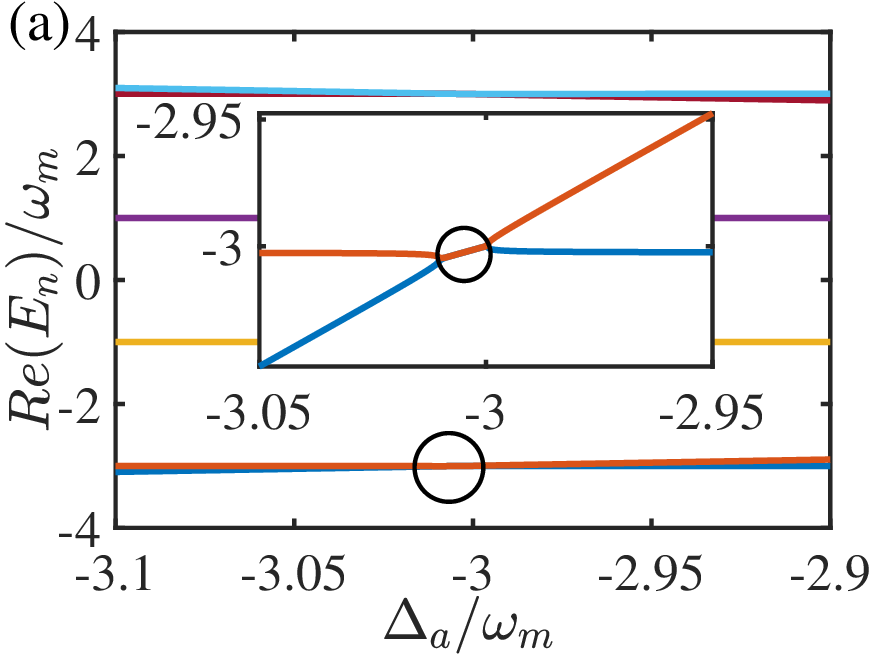}
	\includegraphics[width=0.235\textwidth]{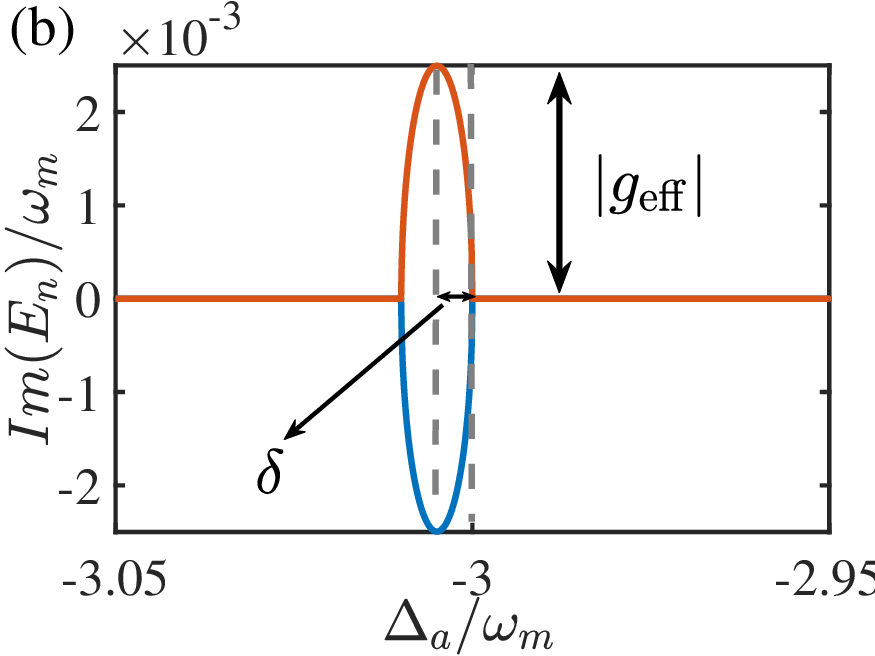}
	\caption{(a) All six real parts of the normalized eigenvalues of the transition matrix $\mathcal{L}$ are plotted as a function of the detuning frequency $\Delta_a/\omega_m$. (b) Two relevant imaginary parts of the normalized eigenvalues are depicted as a function of the detuning frequency $\Delta_a/\omega_m$. The parameters used are $\Delta_b=3\omega_m$ and $g=G=0.1\omega_m$.}\label{eigendiag}
\end{figure}

Dynamics of the system quadrature operators in the Heisenberg picture under the full Hamiltonian in Eq.~\eqref{Hamline} satisfy
\begin{equation}\label{utHeisenberg}
	\dot{u}(t)=i[H,u(t)]=i\mathcal{L}u(t),
\end{equation}
where $u(t)=[X_a(t), Y_a(t), X_b(t), Y_b(t), X_m(t), Y_m(t)]^T$ is the vector of quadrature operators, and $X_m=(m+m^\dag)/\sqrt{2}$, $Y_m=(m-m^\dag)/i\sqrt{2}$. $\mathcal{L}$ represents the whole system's transition matrix, 
\begin{equation}\label{matrix}
\begin{aligned}
\mathcal{L}=i\begin{bmatrix}
	0 &-\Delta_a& 0 & 0 & 0 & 0\\
	\Delta_a& 0 & 0 & 0 & 2g &0\\
	0& 0 & 0 &-\Delta_b & 0 & 0\\
	0& 0 & \Delta_b & 0 & 2G & 0\\
	0& 0 & 0 & 0 & 0 &-\omega_m\\
	2g& 0 & 2G & 0 &\omega_m &0			
\end{bmatrix}.
\end{aligned}
\end{equation}
The Heisenberg equation in Eq.~\eqref{utHeisenberg} can be regarded as a discrete Schr\"{o}dinger equation, where $u(t)$ is conceptualized as an effective operator wave function~\cite{Delocalization}. In the energy-level diagram of the matrix $\mathcal{L}$ as a function of $\Delta_a$, the two-mode squeezing interaction can be illustrated via the level attractions of the real parts and the maximal splittings of the imaginary parts~\cite{kerrmagnon}. 

\begin{figure}[t]
	\centering
	\includegraphics[width=0.235\textwidth]{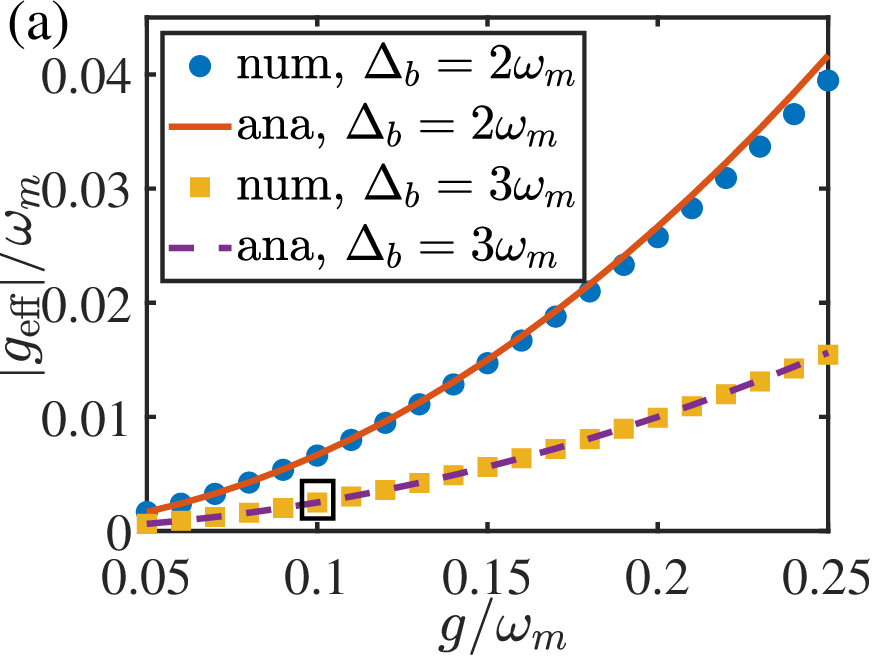}
	\includegraphics[width=0.235\textwidth]{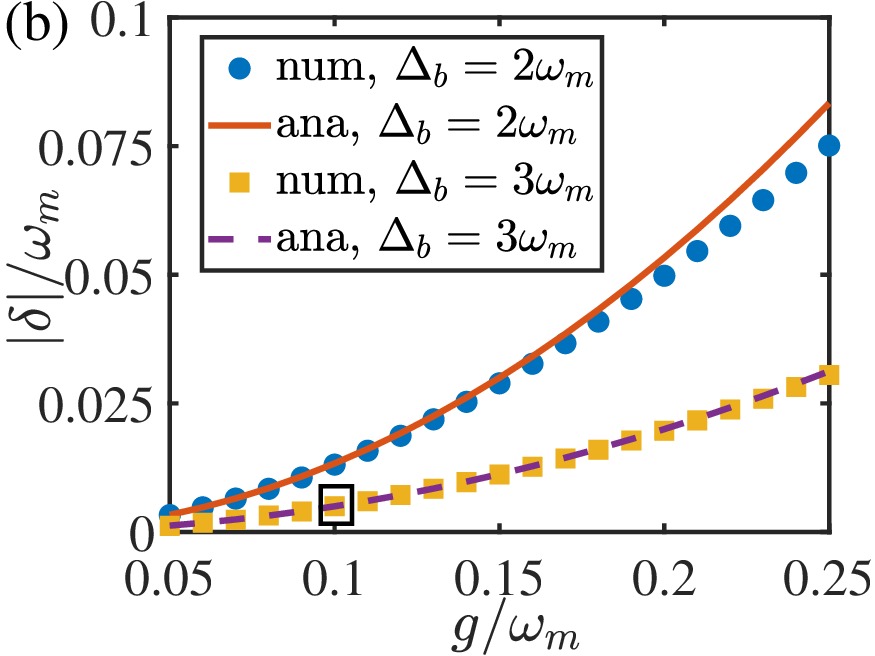}
	\includegraphics[width=0.235\textwidth]{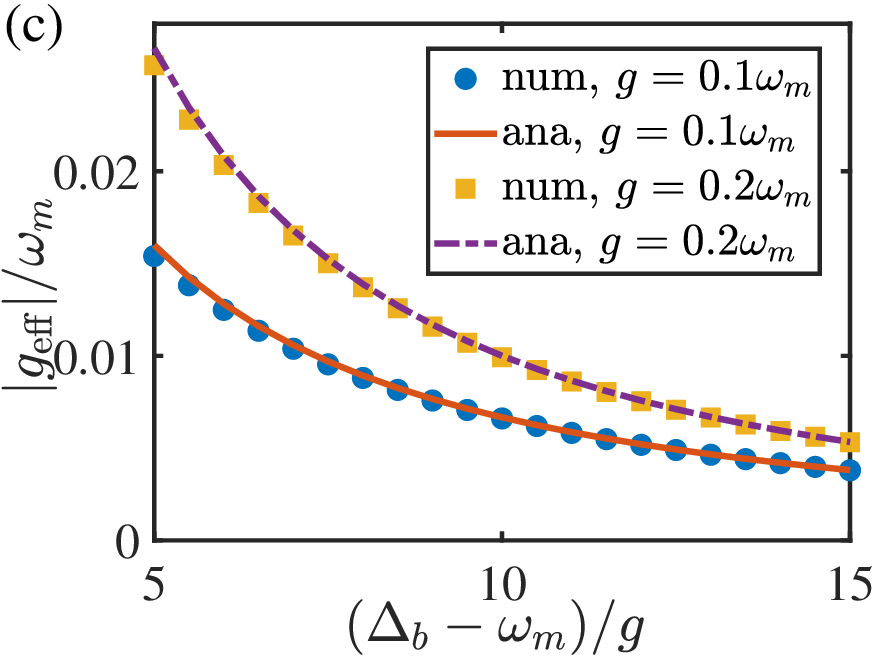}
	\includegraphics[width=0.235\textwidth]{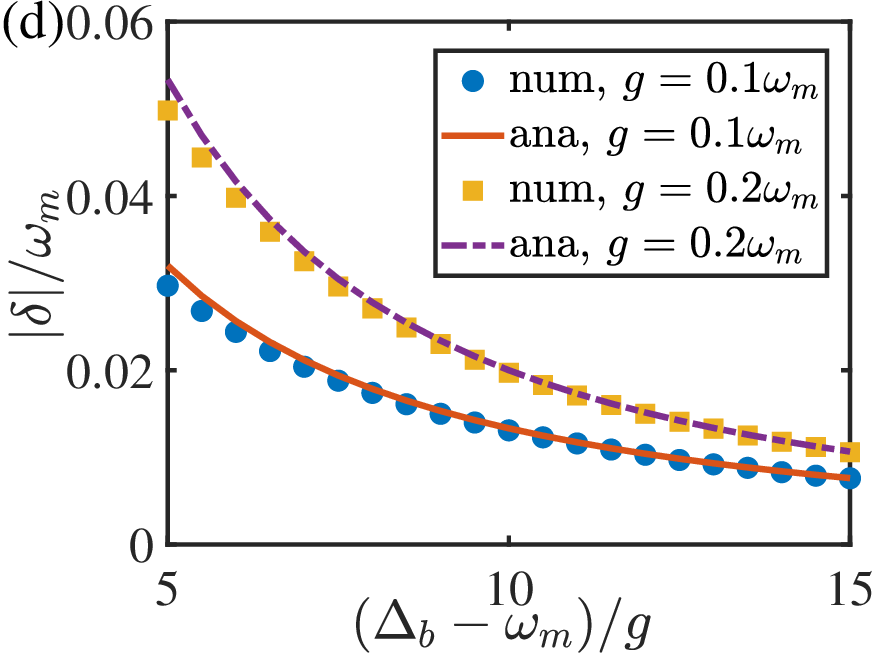}
	\includegraphics[width=0.235\textwidth]{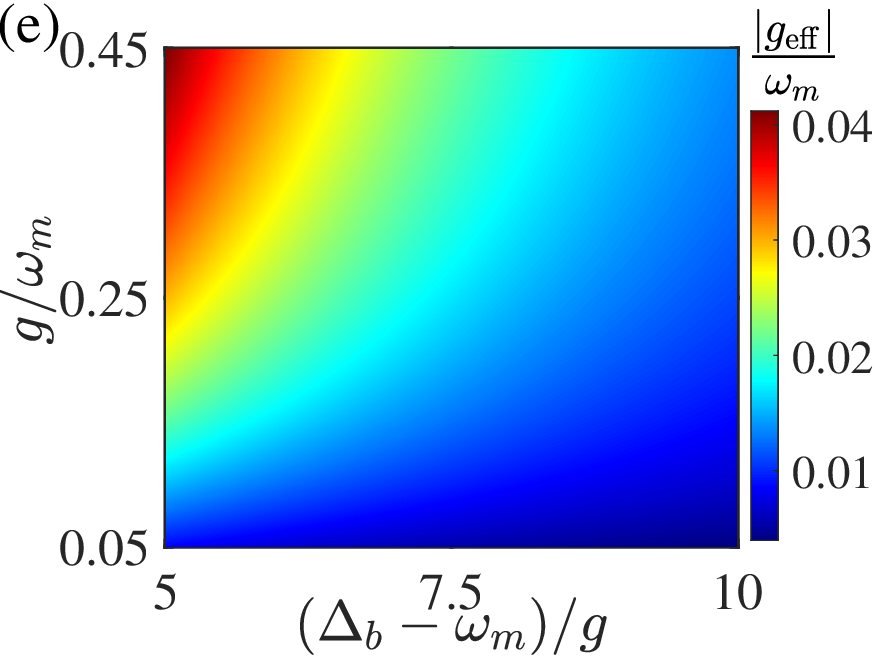}
	\includegraphics[width=0.235\textwidth]{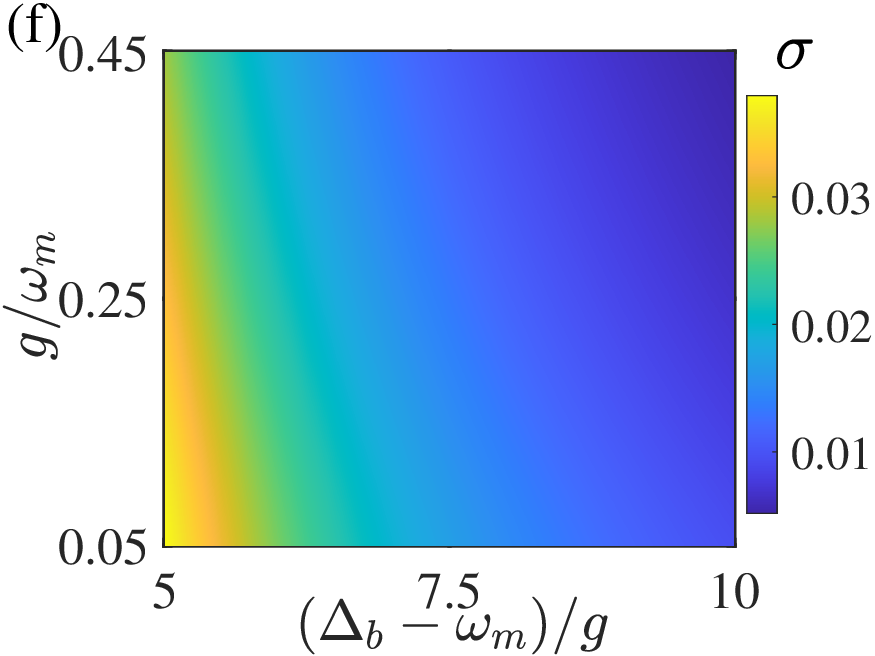}
	\caption{(a) Comparison between the numerically calculated normalized effective coupling strength $|g_{\rm eff}|/\omega_m$ (points) and the corresponding analytical results (lines) in Eq.~\eqref{geffdelta} as a function of $g/\omega_m$. (b) Comparison between the numerically calculated normalized energy shift $|\delta|/\omega_m$ (points) and the corresponding analytical results (lines) in Eq.~\eqref{geffdelta} as a function of $g/\omega_m$. (c) Comparison between the numerically calculated $|g_{\rm eff}|/\omega_m$ (points) and the corresponding analytical results (lines) in Eq.~\eqref{geffdelta} as a function of $(\Delta_b-\omega_m)/g$. (d) Comparison between the numerically calculated $|\delta|/\omega_m$ (points) and the corresponding analytical results (lines) in Eq.~\eqref{geffdelta} as a function of $(\Delta_b-\omega_m)/g$. (e) The numerical effective coupling $|g_{\rm eff}|/\omega_m$ in parameter space spanned by $g$ and $\Delta_b$. (f) The relative deviation $\sigma$ in the parameter space spanned by $g$ and $\Delta_b$. The parameter is fixed as $G=g$.}\label{effgdelta}
\end{figure}

The energy levels of $\mathcal{L}$ in Eq.~\eqref{matrix} are plotted in Figs.~\ref{eigendiag} (a) and (b). The real parts of all six eigenvalues are displayed in Fig.~\ref{eigendiag} (a). The orange and purple lines correspond to the energies of the mechanical mode and are decoupled from the two-mode squeezing dynamics. In contrast, for the remaining four eigenvalues, two distinct energy-level attractions emerge simultaneously as the detuning $\Delta_a$ approaches (but does not exactly equal) $-\Delta_b$. One of the level attractions is highlighted by a dark circle, and the inset further emphasizes it. The imaginary parts of the two relevant eigenvalues (shown in blue and red lines) are presented in Fig.~\ref{eigendiag} (b). As the real parts of the two eigenvalues gradually converge, their imaginary parts grow monotonically, reaching a maximum magnitude of $|g_{\rm eff}|$ at $\Delta_a=-\Delta_b+\delta$. The shift $\delta$ arises from the mutual interaction between the photon and phonon modes.

The maximal splitting $|g_{\rm eff}|$ of the imaginary parts of the two eigenvalues [see Fig.~\ref{eigendiag} (b)] is presented in Fig.~\ref{effgdelta} (a) as a function of the photon-phonon coupling strength $g$, and compared with the analytical prediction given by Eq.~\eqref{geffdelta}. Blue dots and orange squares represent the numerical results at $\Delta_b=2\omega_m$ and $\Delta_b=3\omega_m$, respectively, while the red solid and purple dashed lines correspond to the respective analytical results. One can observe that the analytical $g_{\rm eff}$ do match well with their numerical results for the coupling strength $g\le0.25\omega_m$ at large detuning $\Delta_b=3\omega_m$. At a small detuning $\Delta_b=2\omega_m$, the valid range decreases to $g\le0.18\omega_m$. Notably, both parameter sets enter the ultrastrong coupling regime, defined by $g/\omega_m\ge0.1$. The value indicated by the black box corresponds to those in Fig.~\ref{eigendiag} (b). The energy shift $\delta$ in Eq.~\eqref{geffdelta} is validated in Figure~\ref{effgdelta} (b). It is found that the energy shift $\delta$ is valid when $g\le 0.25\omega_m$ at $\Delta_b=3\omega_m$. As $\Delta_b$ decreases into $2\omega_m$, the valid range turns into $g\le 0.18\omega_m$. Similar results for $g_{\rm eff}$ and $\delta$ are shown in Figs.~\ref{effgdelta} (c) and (d) as functions of the detuning-to-coupling ratio $(\Delta_b-\omega_m)/g$. When $(\Delta_b-\omega_m)/g \ge 8$, the analytical predictions agree well with the numerical results, demonstrating the validity of the effective Hamiltonian in this regime. As the detuning increases, the effective coupling strength $g_{\rm eff}$ gradually decreases. By contrast, reducing the detuning leads to growing deviations between analytical and numerical results, indicating that the large-detuning condition required for the effective Hamiltonian is no longer well satisfied and its validity is correspondingly reduced.

More numerical results of the effective coupling strength $g_{\rm eff}$ are presented in Fig.~\ref{effgdelta} (e). It can be observed that a larger coupling $g$ and a smaller difference $(\Delta_b-\omega_m)$ yield a stronger effective coupling strength $g_{\rm eff}$, which is approximately consistent with the analytical result given by Eq.~\eqref{geffdelta}. To more clearly delineate the valid range of the effective Hamiltonian, we introduce the relative errors of the effective coupling strength, as illustrated in Fig.~\ref{effgdelta} (f). The relative error is defined as
\begin{equation}
	\sigma=\left|\frac{{\rm num}(g_{\rm eff})-{\rm ana}(g_{\rm eff})}{{\rm ana}(g_{\rm eff})}\right|,
\end{equation}
where ${\rm num}(g_{\rm eff})$ describes the numerical result of $g_{\rm eff}$ and ${\rm ana}(g_{\rm eff})$ presents the analytical one. From Fig.~\ref{effgdelta} (f), it can be observed that the relative error $\sigma$ decreases approximately as $\Delta_b-\omega_m$ increases, which is consistent with the conditions required by the perturbation theory. For a fixed and small $\Delta_b$, the $\sigma$ exhibits a significant reduction with increasing $g$, despite this trend seeming to diverge from the results in Fig.~\ref{effgdelta} (c). Indeed, as $g$ increases, the absolute error $|{\rm num}(g_{\rm eff})-{\rm ana}(g_{\rm eff})|$ increases, while the relative error $\sigma$ decreases. By synthesizing the results presented in Figs.~\ref{effgdelta} (a)-(f), a valid and well-performing region is approximately identified where $8g\le(\Delta_b-\omega_m)\le10g, 0.1\omega_m\le g\le 0.2\omega_m$. Within this region, the effective coupling strength $g_{\rm eff}\gtrsim0.01\omega_m$ and the relative error $\sigma\le0.01$, and the absolute error remains small.

\section{Derivation of the optimal quadrature operator}\label{appcm}
This appendix presents the derivation of the quadrature operator $X$ in Eq.~(\ref{quadratureX}). Based on Eq.~(\ref{cmeff}) and under the initial condition $V^{\rm eff}(0)=Diag[N_a+1/2,N_a+1/2,N_b+1/2,N_b+1/2]$, the nonzero matrix elements of the time-dependent CM $V^{\rm eff}(t)$ can be solved as
\begin{equation}\label{Veffelement}
\begin{aligned}
V^{\rm eff}_{11}(t)&=C_+(1-\sin\varphi)e^{(\Omega-\kappa_a-\kappa_b)t}-C_0\cos\varphi e^{-(\kappa_a+\kappa_b)t}\\
&+C_-(1+\sin\varphi)e^{-(\Omega+\kappa_a+\kappa_b)t}+c_a,\\
V^{\rm eff}_{44}(t)&=C_+(1+\sin\varphi)e^{(\Omega-\kappa_a-\kappa_b)t}+C_0\cos\varphi e^{-(\kappa_a+\kappa_b)t}\\
&+C_-(1-\sin\varphi)e^{-(\Omega+\kappa_a+\kappa_b)t}+c_b,\\
V^{\rm eff}_{14}(t)&=-C_+\cos\varphi e^{(\Omega-\kappa_a-\kappa_b)t}+C_0\sin\varphi e^{-(\kappa_a+\kappa_b)t}\\
&+C_-\cos\varphi e^{-(\Omega+\kappa_a+\kappa_b)t}+c,\\
\end{aligned}
\end{equation}
and $V_{22}^{\rm eff}(t)=V^{\rm eff}_{11}(t)$, $V^{\rm eff}_{33}(t)=V^{\rm eff}_{44}(t)$, $V^{\rm eff}_{23}(t)=V^{\rm eff}_{32}(t)=V^{\rm eff}_{41}(t)=V^{\rm eff}_{14}(t)$. The parameters are defined as
\begin{equation}\label{parameter}
\begin{aligned}
\Omega&=\sqrt{4g^2_{\rm eff}+(\kappa_a-\kappa_b)^2},\quad \tan\varphi=\frac{\kappa_a-\kappa_b}{2g_{\rm eff}},\quad\\ C_0&=\frac{\cos\varphi}{2}\left[\frac{\kappa_-}{(\kappa_a+\kappa_b)}-N_-\right],\\
C_{\pm}&=\pm \frac{\kappa_+\mp\sin\varphi\kappa_-}{4[\Omega\mp(\kappa_a+\kappa_b)]}+\frac{N_+\mp\sin\varphi N_-+1}{4},\\ 
N_\pm&=N_a\pm N_b,\quad \kappa_\pm=\kappa_a(2N_a+1)\pm\kappa_b(2N_b+1).\	
\end{aligned}
\end{equation}
And
\begin{equation}\label{steadyV}
	c=\frac{g_{\rm eff}\kappa_a\kappa_b(N_a+N_b+1)}{(g^2_{\rm eff}-\kappa_a\kappa_b)(\kappa_a+\kappa_b)},~c_o=N_o+\frac{1}{2}-\frac{g_{\rm eff}}{\kappa_o}c,~o=a,b
\end{equation}
which are also the solutions of the matrix elements $V^{\rm eff}_{11}$, $V^{\rm eff}_{44}$, and $V^{\rm eff}_{14}$ in the system stability condition, respectively, obtained by setting $\dot{V}^{\rm eff}=0$ in Eq.~(\ref{cmeff}). These steady CM elements are asymptotic values as $t\to\infty$. The system stability regime requires $g^2_{\rm eff}<\kappa_a\kappa_b$. 

Beyond the system stability region, $g^2_{\rm eff}>\kappa_a\kappa_b$, all the CM elements $V^{\rm eff}_{11}(t)$, $V^{\rm eff}_{44}(t)$, and $V^{\rm eff}_{14}(t)$ in Eq.~(\ref{Veffelement}) exhibit exponential divergence due to the exponential factor $\Omega-\kappa_a-\kappa_b>0$. In the specifical situation when $\kappa_a=\kappa_b$, one can demonstrate the elements $V^{\rm eff}_{11}(t)=V^{\rm eff}_{44}(t)$ by Eq.~(\ref{Veffelement}). However, the CM instability does not imply unstable two-mode squeezing. To find a stable TMSS with a higher SL, we define a general two-mode squeezing operator 
\begin{equation}
	X_\phi=\cos\phi X_a+\sin\phi Y_b,
\end{equation} 
where $\phi$ is an angle to optimize. With the CM definition in Eq.~(\ref{Veffdefin}) and its solution in Eq.~(\ref{Veffelement}), the variance of the general quadrature operator $\Delta X_\phi=\langle X_\phi^2\rangle-\langle X_\phi\rangle^2$ can be described as
\begin{equation}\label{Deltaxeff}
\begin{aligned}
\Delta X_\phi(t)&=\cos^2\phi V^{\rm eff}_{11}(t)+\sin^2\phi V^{\rm eff}_{44}(t)+\sin(2\phi)V^{\rm eff}_{14}(t)\\
&=C_+(1-\sin\tilde{\varphi})e^{(\Omega-\kappa_a-\kappa_b)t}-C_0\cos\tilde{\varphi}e^{-(\kappa_a+\kappa_b)t}\\
&+C_-(1+\sin\tilde{\varphi})e^{-(\Omega+\kappa_a+\kappa_b)t}+C_\phi,\\
\end{aligned}
\end{equation}
where $\tilde{\varphi}=\varphi+2\phi$ and $C_\phi=\cos^2\phi c_a+\sin^2\phi c_b+\sin(2\phi)c$, $c_a, c_b, c$ are constants in Eq.~(\ref{steadyV}).

From Eq.~(\ref{Deltaxeff}), one can find that the exponential divergence term in $\Delta X_\phi(t)$ can be canceled when $\tilde{\varphi}=\pi/2$. This implies the existence of an optimized angle $\tilde{\phi}$, which satisfies
\begin{equation}\label{phi}
	\tan(2\tilde{\phi})=\cot(\varphi)=\frac{2g_{\rm eff}}{\kappa_a-\kappa_b}.
\end{equation}
Specifically, $\tilde{\phi}=\pi/4$ at $\kappa_a=\kappa_b$. At this optimized angle $\tilde{\phi}$, the corresponding quadrature operator is given by 
\begin{equation}\label{quadratureXapp}
	X=\cos\tilde{\phi}X_a+\sin\tilde{\phi}Y_b,
\end{equation}
and the associated variance becomes
\begin{equation}\label{Deltaxphitapp}
	\Delta X(t)=2C_-e^{-(\Omega+\kappa_a+\kappa_b)t}+\frac{N_++1+\cos(2\tilde{\phi})N_-}{2}-2C_-,
\end{equation}
which is the same as Eq.~(\ref{Deltaxphit}).

\bibliographystyle{elsarticle-num}  
\bibliography{reference}

@article{quanentangle,
	title = {Quantum entanglement},
	author = {Horodecki, Ryszard and Horodecki, Pawe\l{} and Horodecki, Micha\l{} and Horodecki, Karol},
	journal = {Rev. Mod. Phys.},
	volume = {81},
	issue = {2},
	pages = {865--942},
	numpages = {0},
	year = {2009},
	month = {Jun},
	publisher = {American Physical Society},
	}

@article{quantumcomputing,
  title={Quantum computers},
  author={Ladd, T. D. and Jelezko, F. and Laflamme, R. and Nakamura, Y. and Monroe, C. and O'Brien, J. L.},
  journal={Nature},
  volume={464},
  pages={45},
  year={2010},
  }

@article{quantumsense,
  title = {Quantum sensing},
  author = {Degen, C. L. and Reinhard, F. and Cappellaro, P.},
  journal = {Rev. Mod. Phys.},
  volume = {89},
  issue = {3},
  pages = {035002},
  numpages = {39},
  year = {2017},
  month = {Jul},
  publisher = {American Physical Society},
  }

@article{quantumcommunication,
  title = {Cavity-based quantum networks with single atoms and optical photons},
  author = {Reiserer, Andreas and Rempe, Gerhard},
  journal = {Rev. Mod. Phys.},
  volume = {87},
  issue = {4},
  pages = {1379--1418},
  numpages = {40},
  year = {2015},
  month = {Dec},
  publisher = {American Physical Society},
  }

@article{cavityopto,
	title = {Cavity optomechanics},
	author = {Aspelmeyer, Markus and Kippenberg, Tobias J. and Marquardt, Florian},
	journal = {Rev. Mod. Phys.},
	volume = {86},
	issue = {4},
	pages = {1391--1452},
	numpages = {62},
	year = {2014},
	month = {Dec},
	publisher = {American Physical Society},
	}

@article{magnoncavity,
	title = {Cavity magnonmechanics},
	author = {Zhang, X. and Zou, C.-L. and Jiang, L. and Tang, H.X.},
	journal = {Sci. Adv.},
	volume = {2},
	pages = {e1501286},
	year = {2016},
	}

@article{mppentang,
	title = {Magnon-Photon-Phonon Entanglement in Cavity Magnomechanics},
	author = {Li, Jie and Zhu, Shi-Yao and Agarwal, G. S.},
	journal = {Phys. Rev. Lett.},
	volume = {121},
	issue = {20},
	pages = {203601},
	numpages = {6},
	year = {2018},
	month = {Nov},
	publisher = {American Physical Society},
	}

@article{magnonqubit,
	title = {Entanglement-based single-shot detection of a single magnon with a superconducting qubit},
	author = {Lachance-Quirion, Dany and Piotr Wolski, Samuel and Tabuchi, Yutaka and Kono, Shingo and Usami, Koji and Nakamura, Yasunobu},
	journal = {Science},
	volume = {367},
	pages = {425},
	year = {2020},
	}

@article{Schrodingerstate,
	title={Generation of multicomponent atomic Schr\"{o}dinger cat states of up to 20 qubits},
	author={Song, Chao and Xu, Kai and  Li, Hekang and Zhang, Yu-Ran and Zhang, Xu and Liu, Wuxin and Guo, Qiujiang and Wang, Zhen and Ren, Wenhui and Hao, Jie and Feng, Hui and Fan, Heng and Zheng, Dongning and Wang, Da-Wei and Wang, H. and Zhu, Shi-Yao},
	journal = {Science},
	year={2019},
	volume={365},
	pages={574},
}

@article{ghzstate,
	title = {Generation and Control of Greenberger-Horne-Zeilinger Entanglement in Superconducting Circuits},
	author = {Wei, L. F. and Liu, Y.-x. and Nori, Franco},
	journal = {Phys. Rev. Lett.},
	volume = {96},
	issue = {24},
	pages = {246803},
	numpages = {4},
	year = {2006},
	month = {Jun},
	publisher = {American Physical Society},
	}

@article{superconduct,
	year={2018},
	title={Experimental Greenberger-Horne-Zeilinger entanglement beyond qubits},
	volume={12},
	pages={759},
	author={Erhard, M. and Malik, M. and Krenn, M. and Zeilinger, A.},
	journal={Nat. Photon.}
}

@article{cqed,
	title={Atomic physics and quantum optics using superconducting circuits},
	author={You, J. Q. and Franco, Nori},
	journal={Nature},
	volume={474},
	number={7353},
	pages={589-97},
	year={2011},
	}

@article{noon,
	title = {Generating entangled states from coherent states in circuit QED},
	author = {Qi, Shi-fan and Jing, Jun},
	journal = {Phys. Rev. A},
	volume = {107},
	issue = {4},
	pages = {042412},
	numpages = {12},
	year = {2023},
	month = {Apr},
	publisher = {American Physical Society},
	}

@article{atom,
	title = {Rydberg Atom Entanglements in the Weak Coupling Regime},
	author = {Jo, Hanlae and Song, Yunheung and Kim, Minhyuk and Ahn, Jaewook},
	journal = {Phys. Rev. Lett.},
	volume = {124},
	issue = {3},
	pages = {033603},
	numpages = {5},
	year = {2020},
	month = {Jan},
	publisher = {American Physical Society},
	}

@article{rydberg,
	title = {High-fidelity entanglement and detection of alkaline-earth Rydberg atoms},
	author = {Madjarov, I. S. and Covet, J. P. and Shaw, A. L. and Choi, J. and Kale, A. and Cooper, A. and Pichler, H. and Schkolnik, V. and Williams, J. R. and Endres, M.},
	journal = {Nat. Phys.},
	volume = {16},
	pages = {857},
	year = {2020},
	month = {Aug},
	}

@article{entangledetection,
	title = {Entanglement Detection Length of Multipartite Quantum States},
	author = {Shi, Fei and Chen, Lin and Chiribella, Giulio and Zhao, Qi},
	journal = {Phys. Rev. Lett.},
	volume = {134},
	issue = {5},
	pages = {050201},
	numpages = {6},
	year = {2025},
	month = {Feb},
	publisher = {American Physical Society},	
}

@article{Cvoptical,
	title = {Multipartite continuous-variable optical quantum entanglement: Generation and application},
	author = {Asavanant, Warit and Furusawa, Akira},
	journal = {Phys. Rev. A},
	volume = {109},
	issue = {4},
	pages = {040101},
	numpages = {25},
	year = {2024},
	month = {Apr},
	publisher = {American Physical Society},
	}

@article{Delocalization,
	title = {Quantum Delocalization on Correlation Landscape: The Key to Exponentially Fast Multipartite Entanglement Generation},
	author = {Chu, Yaoming and Li, Xiangbei and Cai, Jianming},
	journal = {Phys. Rev. Lett.},
	volume = {133},
	issue = {11},
	pages = {110201},
	numpages = {7},
	year = {2024},
	month = {Sep},
	publisher = {American Physical Society},
	}

@article{ppnvcenter,
	title = {Photon–phonon entanglement and spin squeezing via dynamically strain-mediated Kerr nonlinearity in dressed nitrogen–vacancy centers},
	journal = {Opt. Laser Technol.},
	volume = {176},
	pages = {110984},
	year = {2024},
	author = {Guanghui Wang and Zhiyuan Li and Xuan Qin and Zhengcai Yang and Xinke Li and Xiao Wu and Yuan Zhou and Yaojia Chen},
}

@article{High,
	title = {Detecting High-Dimensional Entanglement in Cold-Atom Quantum Simulators},
	author = {Euler, Niklas and G\"arttner, Martin},
	journal = {PRX Quantum},
	volume = {4},
	issue = {4},
	pages = {040338},
	numpages = {31},
	year = {2023},
	month = {Dec},
	publisher = {American Physical Society},
	}

@article{quancomcon,
	title = {Quantum Computation over Continuous Variables},
	author = {Lloyd, Seth and Braunstein, Samuel L.},
	journal = {Phys. Rev. Lett.},
	volume = {82},
	issue = {8},
	pages = {1784--1787},
	numpages = {0},
	year = {1999},
	month = {Feb},
	publisher = {American Physical Society},
	}

@article{quaninfcon,
	title = {Quantum information with continuous variables},
	author = {Braunstein, Samuel L. and van Loock, Peter},
	journal = {Rev. Mod. Phys.},
	volume = {77},
	issue = {2},
	pages = {513--577},
	numpages = {0},
	year = {2005},
	month = {Jun},
	publisher = {American Physical Society},
	}

@article{quantelep,
	title={Unconditional quantum teleportation},
	author={Furusawa, A. and S{\o}rensen, J. L. and Braunstein, S. L. and Fuchs, C. A. and Kimble, H. J. and Polzil, E. S.},
	journal={Science},
	volume={282},
	pages={706},
	year={1998},
	}

@article{quantmetro,
	author = {Giovannetti, Vittorio and Lloyd, Seth and Maccone, Lorenzo},
	journal = {Nat. Photon.},
	pages = {222},
	title = {Advanced in quantum metrology},
	volume = {5},
	year = {2011},
	}

@article{convariable,
	title={Continuous-variable entanglement on a chip},
	author={Masada, Genta and Miyata, Kazunori and Politi, Alberto and Hashimoto, Toshikazu and L.O'Brien, Jeremy and Furusawa, Akira},
	journal={Nat. Photon.},
	volume={9},
	pages={316},
	year={2015},
	}

@article{twocolor,
	title = {Generation of Bright Two-Color Continuous Variable Entanglement},
	author = {Villar, A. S. and Cruz, L. S. and Cassemiro, K. N. and Martinelli, M. and Nussenzveig, P.},
	journal = {Phys. Rev. Lett.},
	volume = {95},
	issue = {24},
	pages = {243603},
	numpages = {4},
	year = {2005},
	month = {Dec},
	publisher = {American Physical Society},
	}

@article{twinlaser,
	title = {Observation of Quantum Noise Reduction on Twin Laser Beams},
	author = {Heidmann, A. and Horowicz, R. J. and Reynaud, S. and Giacobino, E. and Fabre, C. and Camy, G.},
	journal = {Phys. Rev. Lett.},
	volume = {59},
	issue = {22},
	pages = {2555--2557},
	numpages = {0},
	year = {1987},
	month = {Nov},
	publisher = {American Physical Society},
	}

@article{stablethreshold,
	title = {Entangling microwaves with light},
	author = {Sahu, Rishabh and Qiu, Liu and Hease, William and Arnold, Georg and Minoguchi, Yuri and Rabl, Peter and Fink, J.M.},
	journal = {Science},
	volume = {380},
	pages = {718},
	year = {2023},
	}

@article{quanphasenpo,
	title = {Quantum Correlations of Phase in Nondegenerate Parametric Oscillation},
	author = {Reid, M. D. and Drummond, P. D.},
	journal = {Phys. Rev. Lett.},
	volume = {60},
	issue = {26},
	pages = {2731--2733},
	numpages = {0},
	year = {1988},
	month = {Jun},
	publisher = {American Physical Society},
	}

@article{realEPR,
	title = {Realization of the Einstein-Podolsky-Rosen paradox for continuous variables},
	author = {Ou, Z. Y. and Pereira, S. F. and Kimble, H. J. and Peng, K. C.},
	journal = {Phys. Rev. Lett.},
	volume = {68},
	issue = {25},
	pages = {3663--3666},
	numpages = {0},
	year = {1992},
	month = {Jun},
	publisher = {American Physical Society},
	}

@article{observasquee,
	title={Observation of squeezed light from one atom excited with two photons},
	author={Qurjoumtsev, A. and Kubanek, A. and Koch, M. and Sames, C. and Pinkse, P. W. H. and Rempe, G. and Murr, K.},
	journal={Nature},
	volume={474},
	pages={623},
	year={2011},
	}

@article{macroscopic,
	title={Experimental long-lived entanglement of two macroscopic objects},
	author={Julsgaard, B. and Kozhekin, A. and Polzik, E. S.},
	journal={Nature},
	volume={413},
	pages={400},
	year={2001},
	}

@article{homodyne,
	title={Atomic homodyne detection of continuous-variable entangled twin-atom states},
	author={Gross, C. and Strobel, H. and Nicklas, E. and Zibold, T. and Bar-Gill, N. and Kurizki, G. and Oberthaler, M. K.},
	journal={Nature},
	volume={480},
	pages={219},
	year={2011},
	}

@article{strongobserspin,
	title = {Strong Quantum Spin Correlations Observed in Atomic Spin Mixing},
	author = {Bookjans, Eva M. and Hamley, Christopher D. and Chapman, Michael S.},
	journal = {Phys. Rev. Lett.},
	volume = {107},
	issue = {21},
	pages = {210406},
	numpages = {5},
	year = {2011},
	month = {Nov},
	publisher = {American Physical Society},
	}

@article{probespin,
	title = {Probing Spin Correlations in a Bose-Einstein Condensate Near the Single-Atom Level},
	author = {Qu, An and Evrard, Bertrand and Dalibard, Jean and Gerbier, Fabrice},
	journal = {Phys. Rev. Lett.},
	volume = {125},
	issue = {3},
	pages = {033401},
	numpages = {6},
	year = {2020},
	month = {Jul},
	publisher = {American Physical Society},
	}

@article{spinBES,
	title = {Emission of Spin-Correlated Matter-Wave Jets from Spinor Bose-Einstein Condensates},
	author = {Kim, Kyungtae and Hur, Junhyeok and Huh, SeungJung and Choi, Soonwon and Choi, Jae-yoon},
	journal = {Phys. Rev. Lett.},
	volume = {127},
	issue = {4},
	pages = {043401},
	numpages = {6},
	year = {2021},
	month = {Jul},
	publisher = {American Physical Society},
	}

@article{phasemeasure,
	title = {Bosonic Pair Production and Squeezing for Optical Phase Measurements in Long-Lived Dipoles Coupled to a Cavity},
	author = {Sundar, Bhuvanesh and Barberena, Diego and Orioli, Asier Pi\~neiro and Chu, Anjun and Thompson, James K. and Rey, Ana Maria and Lewis-Swan, Robert J.},
	journal = {Phys. Rev. Lett.},
	volume = {130},
	issue = {11},
	pages = {113202},
	numpages = {7},
	year = {2023},
	month = {Mar},
	publisher = {American Physical Society},
	}

@article{manipulatinggrowth,
	title = {Manipulating Growth and Propagation of Correlations in Dipolar Multilayers: From Pair Production to Bosonic Kitaev Models},
	author = {Bilitewski, Thomas and Rey, Ana Maria},
	journal = {Phys. Rev. Lett.},
	volume = {131},
	issue = {5},
	pages = {053001},
	numpages = {7},
	year = {2023},
	month = {Aug},
	publisher = {American Physical Society},
	}

@article{powerlawspin,
	title = {Two-mode squeezing in Floquet-engineered power-law interacting spin models},
	author = {Duha, Arman and Bilitewski, Thomas},
	journal = {Phys. Rev. A},
	volume = {109},
	issue = {6},
	pages = {L061304},
	numpages = {6},
	year = {2024},
	month = {Jun},
	publisher = {American Physical Society},
	}

@article{multisqueeatomic,
	title = {Effect of Multi-dressing quantization on three-mode quantum squeezing in Nature Non-Hermitian atomic Ensemble},
	journal = {Opt. Laser Technol.},
	volume = {183},
	pages = {112310},
	year = {2025},
	issn = {0030-3992},
}

@article{antiferromagnetic,
	title = {Quantum sensing of antiferromagnetic magnon two-mode squeezed vacuum},
	author = {R\"omling, Anna-Luisa E. and Kamra, Akashdeep},
	journal = {Phys. Rev. B},
	volume = {109},
	issue = {17},
	pages = {174410},
	numpages = {12},
	year = {2024},
	month = {May},
	publisher = {American Physical Society},
	}

@article{travelingwave,
	title = {Observation of Two-Mode Squeezing in a Traveling Wave Parametric Amplifier},
	author = {Esposito, Martina and Ranadive, Arpit and Planat, Luca and Leger, S\'ebastien and Fraudet, Dorian and Jouanny, Vincent and Buisson, Olivier and Guichard, Wiebke and Naud, C\'ecile and Aumentado, Jos\'e and Lecocq, Florent and Roch, Nicolas},
	journal = {Phys. Rev. Lett.},
	volume = {128},
	issue = {15},
	pages = {153603},
	numpages = {7},
	year = {2022},
	month = {Apr},
	publisher = {American Physical Society},
	}

@article{surfacephonon,
	title = {Squeezing and Multimode Entanglement of Surface Acoustic Wave Phonons},
	author = {Andersson, Gustav and Jolin, Shan W. and Scigliuzzo, Marco and Borgani, Riccardo and Thol\'en, Mats O. and Rivera Hern\'andez, J.C. and Shumeiko, Vitaly and Haviland, David B. and Delsing, Per},
	journal = {PRX Quantum},
	volume = {3},
	issue = {1},
	pages = {010312},
	numpages = {16},
	year = {2022},
	month = {Jan},
	publisher = {American Physical Society},
	}

@article{nonlocalryd,
	title = {Nonlocal Rydberg enhancement for four-wave-mixing biphoton generation},
	author = {Zhao, Hui-Min and Zhang, Xiao-Jun and Artoni, M. and La Rocca, G. C. and Wu, Jin-Hui},
	journal = {Phys. Rev. A},
	volume = {109},
	issue = {4},
	pages = {043711},
	numpages = {11},
	year = {2024},
	month = {Apr},
	publisher = {American Physical Society},
	}

@article{Nonrecisteer,
	title = {Nonreciprocal steering between optical and microwave waves by Bogoliubov cooling in a cavity optomagnonic system},
	author = {Kong, Deyi and Wang, Fei},
	journal = {Phys. Rev. A},
	volume = {111},
	issue = {1},
	pages = {013704},
	numpages = {12},
	year = {2025},
	month = {Jan},
	publisher = {American Physical Society},
	}

@article{kerrmagnon,
	title = {Kerr-magnon-assisted asymptotic stationary photon-phonon squeezing},
	author = {Qi, Shi-fan and Jing, Jun},
	journal = {Phys. Rev. A},
	volume = {111},
	issue = {1},
	pages = {013708},
	numpages = {12},
	year = {2025},
	month = {Jan},
	publisher = {American Physical Society},
	}

@article{Optoacoustic,
	title = {Optoacoustic Entanglement in a Continuous Brillouin-Active Solid State System},
	author = {Zhu, Changlong and Genes, Claudiu and Stiller, Birgit},
	journal = {Phys. Rev. Lett.},
	volume = {133},
	issue = {20},
	pages = {203602},
	numpages = {7},
	year = {2024},
	month = {Nov},
	publisher = {American Physical Society},
	}

@article{reservoiroptome,
	title = {Reservoir-Engineered Entanglement in Optomechanical Systems},
	author = {Wang, Ying-Dan and Clerk, Aashish A.},
	journal = {Phys. Rev. Lett.},
	volume = {110},
	issue = {25},
	pages = {253601},
	numpages = {5},
	year = {2013},
	month = {Jun},
	publisher = {American Physical Society},
	}

@article{entanglemacro,
	title = {Entangling Macroscopic Oscillators Exploiting Radiation Pressure},
	author = {Mancini, Stefano and Giovannetti, Vittorio and Vitali, David and Tombesi, Paolo},
	journal = {Phys. Rev. Lett.},
	volume = {88},
	issue = {12},
	pages = {120401},
	numpages = {4},
	year = {2002},
	month = {Mar},
	publisher = {American Physical Society},
	}

@article{probeentangle,
	title = {Creating and Probing Multipartite Macroscopic Entanglement with Light},
	author = {Paternostro, M. and Vitali, D. and Gigan, S. and Kim, M. S. and Brukner, C. and Eisert, J. and Aspelmeyer, M.},
	journal = {Phys. Rev. Lett.},
	volume = {99},
	issue = {25},
	pages = {250401},
	numpages = {4},
	year = {2007},
	month = {Dec},
	publisher = {American Physical Society},
	}

@article{Gauinf,
	title = {Gaussian quantum information},
	author = {Weedbrook, Christian and Pirandola, Stefano and Garc\'{\i}a-Patr\'on, Ra\'ul and Cerf, Nicolas J. and Ralph, Timothy C. and Shapiro, Jeffrey H. and Lloyd, Seth},
	journal = {Rev. Mod. Phys.},
	volume = {84},
	issue = {2},
	pages = {621--669},
	numpages = {0},
	year = {2012},
	month = {May},
	publisher = {American Physical Society},
	}

@article{robustopto,
	title = {Robust Photon Entanglement via Quantum Interference in Optomechanical Interfaces},
	author = {Tian, Lin},
	journal = {Phys. Rev. Lett.},
	volume = {110},
	issue = {23},
	pages = {233602},
	numpages = {5},
	year = {2013},
	month = {Jun},
	publisher = {American Physical Society},
	}

@article{doubleoptosys,
	title = {Generation of broadband two-mode squeezed light in cascaded double-cavity optomechanical systems},
	author = {Li, Zhen and Ma, Sheng-li and Li, Fu-li},
	journal = {Phys. Rev. A},
	volume = {92},
	issue = {2},
	pages = {023856},
	numpages = {10},
	year = {2015},
	month = {Aug},
	publisher = {American Physical Society},
	}

@article{thermalsquee,
	title = {Dynamical Two-Mode Squeezing of Thermal Fluctuations in a Cavity Optomechanical System},
	author = {Pontin, A. and Bonaldi, M. and Borrielli, A. and Marconi, L. and Marino, F. and Pandraud, G. and Prodi, G. A. and Sarro, P. M. and Serra, E. and Marin, F.},
	journal = {Phys. Rev. Lett.},
	volume = {116},
	issue = {10},
	pages = {103601},
	numpages = {6},
	year = {2016},
	month = {Mar},
	publisher = {American Physical Society},
	}

@article{twomechansquee,
	title = {Dissipation-driven two-mode mechanical squeezed states in optomechanical systems},
	author = {Tan, Huatang and Li, Gaoxiang and Meystre, P.},
	journal = {Phys. Rev. A},
	volume = {87},
	issue = {3},
	pages = {033829},
	numpages = {7},
	year = {2013},
	month = {Mar},
	publisher = {American Physical Society},
	}

@article{intermagnon,
	title = {Magnon-assisted photon-phonon conversion in the presence of structured environments},
	author = {Qi, Shi-fan and Jing, Jun},
	journal = {Phys. Rev. A},
	volume = {103},
	issue = {4},
	pages = {043704},
	numpages = {10},
	year = {2021},
	month = {Apr},
	publisher = {American Physical Society},
	}

@article{james,
	title = {Generalized James' effective Hamiltonian method},
	author = {Shao, Wenjun and Wu, Chunfeng and Feng, Xun-Li},
	journal = {Phys. Rev. A},
	volume = {95},
	issue = {3},
	pages = {032124},
	numpages = {4},
	year = {2017},
	month = {Mar},
	publisher = {American Physical Society},
	}

@article{oneexcite,
	title = {One Photon Can Simultaneously Excite Two or More Atoms},
	author = {Garziano, Luigi and Macr\`{\i}, Vincenzo and Stassi, Roberto and Di Stefano, Omar and Nori, Franco and Savasta, Salvatore},
	journal = {Phys. Rev. Lett.},
	volume = {117},
	issue = {4},
	pages = {043601},
	numpages = {6},
	year = {2016},
	month = {Jul},
	publisher = {American Physical Society},
	}

@article{nonlinear,
	title = {Deterministic quantum nonlinear optics with single atoms and virtual photons},
	author = {Kockum, Anton Frisk and Miranowicz, Adam and Macr\`{\i}, Vincenzo and Savasta, Salvatore and Nori, Franco},
	journal = {Phys. Rev. A},
	volume = {95},
	issue = {6},
	pages = {063849},
	numpages = {23},
	year = {2017},
	month = {Jun},
	publisher = {American Physical Society},
	}

@article{Entangling,
	title = {Entangling two microwave modes via optomechanics},
	author = {Cai, Qizhi and Liao, Jinkun and Zhou, Qiang},
	journal = {Phys. Rev. A},
	volume = {100},
	issue = {4},
	pages = {042330},
	numpages = {8},
	year = {2019},
	month = {Oct},
	publisher = {American Physical Society},
	}

@article{nonreciprocal,
	title = {Nonreciprocal entanglement in cavity-magnon optomechanics},
	author = {Chen, Jiaojiao and Fan, Xiao-Gang and Xiong, Wei and Wang, Dong and Ye, Liu},
	journal = {Phys. Rev. B},
	volume = {108},
	issue = {2},
	pages = {024105},
	numpages = {8},
	year = {2023},
	month = {Jul},
	publisher = {American Physical Society},
	}

@article{generobust,
	title={Generation of robust optical entanglement on cavity optomagnonics},
	author={Xie, Hong and He, Le-Wei and Liao, Chang-Geng and Chen, Zhi-Hua and Lin, Xiu-Min},
	journal={Opt. Express},
	volume={31},
	pages={7994},
	year={2023},
	}

@article{synergizing,
	title = {Enhancing tripartite photon-phonon-magnon entanglement by synergizing parametric amplifications},
	author = {Wang, Yan and Wu, Jin-Lei and Jiao, Ya-Feng and Lu, Tian-Xiang and Zhang, Hui-Lai and Jiang, Li-Ying and Kuang, Le-Man and Jing, Hui},
	journal = {Phys. Rev. A},
	volume = {111},
	issue = {1},
	pages = {013709},
	numpages = {11},
	year = {2025},
	month = {Jan},
	publisher = {American Physical Society},
	}

@article{Xie2023,
	year = {2023},
	month = {may},
	publisher = {IOP Publishing},
	volume = {8},
	number = {3},
	pages = {035022},
	author = {Xie, Jikun and Yuan, Huaiyang and Ma, Shengli and Gao, Shaoyan and Li, Fuli and Duine, Rembert A},
	title = {Stationary quantum entanglement and steering between two distant macromagnets},
	journal = {Quantum Sci. Technol.},
}

@article{cohermechanical,
   title={Coherent state transfer between itinerant microwave fields and a mechanical oscillator},
   author={Palomakim, T. A. and Harlow, J. W. and Teufel, J. D. and Simmonds, R. W. and Lehnert, K. W.},
   year={2013},
   journal={Nature},
   volume={495},
   pages={210},
   }

@article{entmechanical,
	title={Entangling mechanical motion with microwave fields},
	author={Palomakim, T. A. and Teufel, J. D. and Simmonds, R. W. and Lehnert, K. W.},
	year={2013},
	journal={Science},
	volume={342},
	pages={710},
	}

@article{Atomicoscill,
	title = {Creation of Two-Mode Squeezed States in Atomic Mechanical Oscillators},
	author = {Leong, Wui Seng and Xin, Mingjie and Chen, Zilong and Wang, Yu and Lan, Shau-Yu},
	journal = {Phys. Rev. Lett.},
	volume = {131},
	issue = {19},
	pages = {193601},
	numpages = {5},
	year = {2023},
	month = {Nov},
	publisher = {American Physical Society},
	}

@article{squeemech,
	title={Quantum squeezing of motion in a mechanical resonator},
	author={Wollman, E. E. and Lei, C. U. and Weinstein, A. J. and Suh, J. and Kronwald, A. and Marquardt, F. and Clerk, A. A. and Schwab, K. C.},
	year={2015},
	journal={Science},
	volume={349},
	pages={952},
	}

@article{map,
	title = {Mapping the Cavity Optomechanical Interaction with Subwavelength-Sized Ultrasensitive Nanomechanical Force Sensors},
	author = {Fogliano, Francesco and Besga, Benjamin and Reigue, Antoine and Heringlake, Philip and Mercier de L\'epinay, Laure and Vaneph, Cyril and Reichel, Jakob and Pigeau, Benjamin and Arcizet, Olivier},
	journal = {Phys. Rev. X},
	volume = {11},
	issue = {2},
	pages = {021009},
	numpages = {20},
	year = {2021},
	month = {Apr},
	publisher = {American Physical Society},
	}

@article{Superposition,
	title = {Macroscopic Quantum Superposition in Cavity Optomechanics},
	author = {Liao, Jie-Qiao and Tian, Lin},
	journal = {Phys. Rev. Lett.},
	volume = {116},
	issue = {16},
	pages = {163602},
	numpages = {6},
	year = {2016},
	month = {Apr},
	publisher = {American Physical Society},
	}

@article{coherentcoupling,
	title = {Coherent Coupling between Phonons, Magnons, and Photons},
	author = {Shen, Zhen and Xu, Guan-Ting and Zhang, Mai and Zhang, Yan-Lei and Wang, Yu and Chai, Cheng-Zhe and Zou, Chang-Ling and Guo, Guang-Can and Dong, Chun-Hua},
	journal = {Phys. Rev. Lett.},
	volume = {129},
	issue = {24},
	pages = {243601},
	numpages = {7},
	year = {2022},
	month = {Dec},
	publisher = {American Physical Society},
	}

@article{active,
	title={Active optomechanics},
	author={Yu, Deshui and Vollmer Frank},
	year={2022},
	journal={Commun. Phys.},
	volume={5},
	pages={61},
	}

@article{MechanicalBis,
	title = {Mechanical Bistability in Kerr-modified Cavity Magnomechanics},
	author = {Shen, Rui-Chang and Li, Jie and Fan, Zhi-Yuan and Wang, Yi-Pu and You, J. Q.},
	journal = {Phys. Rev. Lett.},
	volume = {129},
	issue = {12},
	pages = {123601},
	numpages = {7},
	year = {2022},
	month = {Sep},
	publisher = {American Physical Society},
	}

@article{Rempe:92,
	author = {G. Rempe and R. J. Thompson and H. J. Kimble and R. Lalezari},
	journal = {Opt. Lett.},
	number = {5},
	pages = {363--365},
	publisher = {Optica Publishing Group},
	title = {Measurement of ultralow losses in an optical interferometer},
	volume = {17},
	month = {Mar},
	year = {1992},
	}

@article{Gorodetsky:96,
	author = {M. L. Gorodetsky and A. A. Savchenkov and V. S. Ilchenko},
	journal = {Opt. Lett.},
	number = {7},
	pages = {453--455},
	publisher = {Optica Publishing Group},
	title = {Ultimate Q of optical microsphere resonators},
	volume = {21},
	month = {Apr},
	year = {1996},
	}

@article{entdynhightem,
	title = {Entangling Two Macroscopic Mechanical Resonators at High Temperature},
	author = {Lin, Qing and He, Bing and Xiao, Min},
	journal = {Phys. Rev. Appl.},
	volume = {13},
	issue = {3},
	pages = {034030},
	numpages = {16},
	year = {2020},
	month = {Mar},
	publisher = {American Physical Society},
}

@article{Chen:17,
	author = {Zhi Xin Chen and Qing Lin and Bing He and Zhi Yang Lin},
	journal = {Opt. Express},
	number = {15},
	pages = {17237--17248},
	publisher = {Optica Publishing Group},
	title = {Entanglement dynamics in double-cavity optomechanical systems},
	volume = {25},
	year = {2017},
}

@article{Xie:20,
	author = {Yi Fei Xie and Zhen Cao and Bing He and Qing Lin},
	journal = {Opt. Express},
	number = {15},
	pages = {22580--22593},
	publisher = {Optica Publishing Group},
	title = {PT-symmetric phonon laser under gain saturation effect},
	volume = {28},
	year = {2020},
}

@ARTICLE{amptmsnanomechanical,
	author={Wodedo, Muhdin Abdo and Tesfahannes, Tesfay Gebremariam and Darge, Tewodros Yirgashewa and Pereira, Mauro and Teklu, Berihu},
	journal={IEEE Trans. Quantum Eng.}, 
	title={Amplifying Two-Mode Squeezing in Nanomechanical Resonators}, 
	year={2025},
	volume={6},
	pages={1-15},
}

@article{WODEDO2025108364,
	title = {Optimizing mechanical entanglement using squeezing and parametric amplification},
	journal = {Results Phys.},
	volume = {76},
	pages = {108364},
	year = {2025},
	issn = {2211-3797},
	author = {Muhdin Abdo Wodedo and Tesfay Gebremariam Tesfahannes and Tewodros Yirgashewa Darge and Berihu Teklu},
}

@article{Frequencyest,
	title={Frequency estimation by frequency jumps},
	journal={arxiv:},
	volume={2503},
	pages={06738},
	year={2025},
	author={Simone Cavazzoni and Berihu Teklu and Matteo G. A. Paris},
}

@article{SCHNABEL20171,
	title = {Squeezed states of light and their applications in laser interferometers},
	journal = {Phys. Rep.},
	volume = {684},
	pages = {1-51},
	year = {2017},
	note = {Squeezed states of light and their applications in laser interferometers},
	issn = {0370-1573},
	author = {Roman Schnabel},
}

@article{Enhancementoem,
	title = {Enhancement of opto-electro-mechanical entanglement through three-level atoms},
	journal = {Phys. Lett. A},
	volume = {525},
	pages = {129920},
	year = {2024},
	issn = {0375-9601},
	author = {Abebe Senbeto Kussia and Tewodros Yirgashewa Darge and Tesfay Gebremariam Tesfahannes and Abeba Teklie Bimeraw and Berihu Teklu},
}
\end{document}